\UseRawInputEncoding
\documentclass[reqno,12pt]{article}
\usepackage{amsmath,amsfonts,amssymb,amsthm,amstext,amscd,eucal,xcolor}
\usepackage[all]{xy}
\usepackage{latexsym}
\usepackage{hyperref}
\usepackage{epsfig}
\usepackage{color}
\usepackage{graphicx}
\usepackage[active]{srcltx}
\usepackage{array}
\usepackage{float}
\usepackage{xcolor}

\makeatletter \@addtoreset{equation}{section}

\makeatletter\renewcommand\section{\@startsection {section}{1}{\z@}%
	{-3.5ex \@plus -1ex \@minus -.2ex}
	{2.3ex \@plus.2ex}%
	{\normalfont\large\bfseries}}
\renewcommand\subsection{\@startsection{subsection}{2}{\z@}%
	{-3.25ex\@plus -1ex \@minus -.2ex}%
	{1.5ex \@plus .2ex}%
	{\normalfont\bfseries}}

\newcommand{\be}{\begin{equation}}
	\newcommand{\ee}{\end{equation}}
\newcommand{\bea}{\begin{eqnarray}}
	\newcommand{\eea}{\end{eqnarray}}
\newcommand{\bse}{\begin{subequations}}
	\newcommand{\ese}{\end{subequations}}
\newcommand{\bi}{\begin{itemize}}
	\newcommand{\ei}{\end{itemize}}

\newcommand{\beq}{\begin{eqnarray}}
	\newcommand{\eeq}{\end{eqnarray}}

\def\s2s1{S$^2\times$S$^1$ }
\def\Label#1{\label{#1}%
	\smash{\hbox to0pt{\raise1ex\hbox{\tiny[#1]}\hss}}}
\def\noLabels{\let\Label=\label}
\def\nobbibitem{\let\bbibitem=\bibitem}

\newcommand*{\rom}[1]{\expandafter\@slowromancap\romannumeral #1@}
\begin{document}
	\baselineskip 18pt%
	\begin{titlepage}
		\vspace*{20mm}
		\begin{center}
			{\Large{\textbf{The effect of redshift function on the Weak Energy Conditions in f(R) Wormholes}}}
			\vspace*{8mm}
			
			Amid Sadeghi Nezhad\footnote{amidsadeghinezhad@phy.uk.ac.ir} \\
			Mohammad Reza Mehdizadeh\footnote{mehdizadeh.mr@uk.ac.ir} \\
			Hanif Golchin\footnote{h.golchin@uk.ac.ir} \\

			\vspace*{0.4cm}
			{ \it Faculty of Physics, Shahid Bahonar University of Kerman, \\
				PO Box 76175, Kerman, Iran} and \\
		
			\vspace*{1.5cm}
		\end{center}

\begin{abstract}
In the present paper, we investigate traversable wormhole solutions determined by an exponential shape function and fractional redshift
function in the background of four viable $f(R)$ models. Although in the absence of the redshift function $\varphi(r)$ the null energy condition (NEC) and weak energy condition (WEC) are violated, we find that considering the redshift function, NEC and WEC are respected by choosing the appropriate parameters in the models. We also investigate the conditions of stability and absence of anti-gravity effects for these wormholes. Our results show that in the case of $\varphi(r) \neq 0$ these conditions are satisfied easier than the case of $\varphi(r)=0$. Finally  we calculate the deflection angle using the gravitational lensing effect. We show that the deflection angle increases by inserting the redshift function. 
\end{abstract}


\end{titlepage}

\addtocontents{toc}{\protect\setcounter{tocdepth}{2}}
\tableofcontents

\section{Introduction}
Wormholes are assumptive topological tunnels which connect two different space-times of the same universe, or  even two separate universes by a minimal surface called the "throat" of the wormhole. Concept of inter-universe connections can be traced back to 1916 through the pioneering work of Flamm \cite{flamm}, shortly after the inception of General Relativity. In 1935, Einstein and Rosen developed a bridge model \cite{EinstenRosen} by using the Flamm’s theories. The notion of a "wormhole" as a concise topological pathway in the space-time was initially postulated by Misner and Wheeler \cite{misner-wheeler}.
In 1988, Lorentzian wormholes, which provide the possibility of bidirectional transit of matter and energy, were initially explored by Morris and Thorne \cite{MT metric} in the framework of Einstein's theory of general relativity.

To assure the dual-directional traversability of the wormhole, the throat's geometry should satisfy the flaring-out condition \cite{misner-wheeler}. This condition ensures that the throat expands as one moves away from its center, preventing collapse and allowing for bidirectional transit of matter and information.
In four-dimensional general relativity, the flaring-out condition leads to the violation of NEC and WEC. This necessitate the presence of "exotic matter" in the structure of the wormhole throat, characterized by a negative energy density \cite{ExM Re}. Despite the fact that exotic matter has not been found in the physical world, efforts are being made to search for evidence of its existence in the cosmos \cite{Ex evidence}. 
There are some investigations with the aim of eliminating or minimizing the use of exotic matter in wormholes construction, by applying of the modified theories of gravity \cite{Visser}. Modified gravity theories  are alternative approaches to rectify the deficiencies or limitations in Einstein's general relativity. In recent years, large number of modified gravity models are investigated in the field of gravity and cosmology \cite{modi grav}. These innovative theories present additional terms to modify the field equations, or to change the geometric structure of the space-time which provide explanations for the physical and cosmological phenomena. For instance, by introducing higher-order terms in curvature, it becomes possible to create thin-shell wormholes that are created by standard matter \cite{thin shell}.

Recently, through the application of modified gravity theories, broad exploration has been done to construct the wormhole configurations that do not rely on exotic matter \cite{harko2013}. In 2007, an investigation of thin-shell wormholes in the context of Einstein-Gauss-Bonnet gravity was carried out \cite{Richarte}. In 2011, descriptions regarding the energy flux emitted onto wormholes in Brans-Dicke theory were provided for Observational Manifestations \cite{Alexeyev 2011}. Properties and existence of the wormhole throats in f(R) gravity is explored  in \cite{Horvat}. Wormhole geometries in the framework of third-order Love-lock gravity is also studied in 2016 \cite{Mehdizadeh Lovelock 2016}. 
In 2016 Lorentzian wormhole solutions are obtained in scalar-tensor gravity \cite{Kar2016} and charged wormhole solutions satisfying NEC and WEC, are derived in Einstein-Cartan gravity \cite{Mehdizadeh and Ziaie Einstein-Cartan} in 2019.
In the framework of cubic gravity, the wormhole solutions that respect the energy conditions at the throat, are found in \cite{Mehdizadeh and Ziaie cubic}.

One of the most widely used modifications to General Relativity is the $f(R)$ gravity which was proposed by Buchadahl \cite{buchadahl} in 1977. In $f(R)$ gravity, the curvature scalar or Ricci scalar $R$ in the Einstein-Hilbert action is replaced by $f(R)$, which represents an arbitrary function of the Ricci scalar $R$. This modification leads to a generalized form of the Einstein field equations in this theory \cite{Pavlovic Sossich 2015}. Indeed, these generalized equations have demonstrated hopeful descriptions of cosmic phenomena. In 2011, the hydrostatic equilibrium of stellar structure was studied in the context of the modified $f(R)$ gravity \cite{Capozziello}. Neutron stars in $f(R)$ gravity were investigated in 2016 \cite{Bakirova}. An analysis of extended stellar kinematics of elliptical galaxies considering the modified $f(R)$ gravity was provided in 2012 \cite{Napolitano}. The gravitational interactions of galaxy clusters were described in 2014, utilizing $f(R)$ gravity \cite{Terukina}. The structure formation and evolution of the universe are also investigated by considering the $f(R)$ gravity effects\cite{Cosmological Evolution in f(R),Planck 2015 results,Arnold2015,Cosmological evolutions in Tsujikawa model of f(R) Gravity}. Traversable wormhole geometries satisfying the energy conditions were achieved by Lobo and Oliveira in 2009 \cite{Wh geometries in f(R) modified theories of gravity}. Lorentzian wormhole solutions were explored by Pavlovica and Sossich in 2015, satisfy the WEC \cite{Pavlovic Sossich 2015}. A class of thin-shell wormhole solutions in the context of $f(R)$ gravity was constructed in 2016\cite{Eiroa 2016}. 
Several viable f(R) models were investigated in 2019 to derive wormhole solutions that satisfy the criteria of the NEC and WEC \cite{Quasi}. 
Another class of wormhole solutions has been investigated in studies \cite{mishra and sharma 2021} \cite{wormholes with constant and variable redshift functions} through the selection of the shape function and redshift function. Bronnikov et al also discussed the wormhole non-existence conditions in the context of f(R) theories \cite{Bronnikov:2010tt}. \\
In this research, our target is to study the effect of redshift function $\varphi(r)$ on wormhole solutions within 4 $f(R)$ gravity models. We are seeking solutions that satisfy the WEC, avoid being anti-gravity cases, and steer clear of cosmic instability. To achieve this, in section ~\ref{s2}, we began by deriving the field equations in context of $f(R)$ gravity for Morris-Thorne metric. Then, in section ~\ref{s3} , we introduced the general forms of our models along with the shape function and the redshift function. We calculated and rewrite the field equations for each model.
Continuing into section ~\ref{s3}, we found wormhole solutions for each model. In section ~\ref{s4}, using the gravitational lensing effect as an observational evident we calculated the effective potential function and by analyzing the deflection angle, we investigated the geodesics around the wormhole.
\section{Field equations and wormhole geometry in f(R) gravity} \label{s2}
We start with the action of modified f(R) gravity as:
\be \label{action}
S=\frac{1}{2\kappa}\int d^4x\sqrt{-g}\,(f(R)+{\cal L}_{M})
\ee
Where ${\cal L}_{M}$  is the Lagrangian of the matter field, $\kappa=8 \pi G$ is the gravitational coupling constant and $g$ is determinant of the metric. In this study, for notational simplicity, we consider $\kappa=1$. The field equation is derived by varying the action (\ref{action}) with respect to metric $g_{\mu\nu}$, resulting in the following fourth-order field equations \cite{Antonio-Tsujikawa}:
\be \label{feild}
	R_{\mu \nu}f_{R}(R) - \frac{1}{2}
	g_{\mu \nu}f(R)-(\nabla_{\mu}\nabla_{\nu}-
	g_{\mu \nu}\Box)f_{R}(R) = k T_{\mu \nu} 
\ee
Where $f_{R}=df(R)/dR$, and $T_{\mu \nu}$ is the energy-momentum tensor that given by varying the matter action $S_{M}$ with respect to metric $g^{\mu\nu}$:
\be \label{EM Tensor}
T_{\mu \nu} = \frac{-2}{\sqrt{-g}} \frac{\delta S_{M}}
{\delta g^{\mu \nu}}
\ee

Now, based on the gravitational field equation $G_{\mu \nu}=R_{\mu \nu}-\frac{1}{2} R g_{\mu \nu}$, we can derive the following equation for Einstein tensor:
\be \label{Einstein tensor}
G_{\mu \nu}=\frac{1}{f_{R}}\lbrace f_{RR}\nabla_{\mu}\nabla_{\nu}R + f_{RRR}(\nabla_{\mu}R)(\nabla_{\nu}R)  \\ - \frac{g_{\mu \nu}}{6}(R f_{R}+ f + 16 \pi GT)+ 8 \pi G T_{\mu \nu}\rbrace
\ee
Here, $T$ represents the trace of the energy-momentum tensor, and $f_{RR}$ and $f_{RRR}$ denote the second and third derivatives of $f(R)$ with respect to $R$. To have cosmologically viable $f(R)$ gravity models, the following conditions for $f_{R}$ and $f_{RR}$ need to be satisfied in region $R>R_{0}$ which $R_0$ represents the Ricci scalar value at the present background curvature \cite{Amendola Dark Energy}:

$a)$ $f_{R}>0$, this is necessary to prevent the occurrence of anti-gravity effects. Note that in this case the wormhole existence condition is violated \cite{Bronnikov:2010tt}, however these solutions are stable against spherically symmetric perturbations and provides an interesting possibility to construct non-static or thin-shell wormhole solutions \cite{Bronnikov:2009tv}
 
$b)$ $f_{RR}>0$, to be consistent with local gravity constraints \cite{Dolgove fRR} and to maintain the stability of cosmological perturbations \cite{Cosmic  perturbations in f(R) gravity}.
 
$c)$ $f_{RR}\ne0$, to prevent the weak singularities which means the divergency in the physical quantities as density and curvature\cite{star07}

In the following, we use the Morris-Thorne proposed metric which describes the 
static and spherically symmetric traversable wormholes \cite{MT metric}.
\be \label{wmetric}
ds^2=-e^{-2\phi(r)}dt^2+\frac{dr^2}{1-b(r)/r}+r^2\,(d\theta^2 +\sin
^2{\theta} \, d\phi ^2) \,.
\ee
In this metric, the functions $b(r)$ and $\phi(r)$ are introduced as arbitrary functions of the embedding-space radial coordinate $r$, which are called the shape function and the redshift function, respectively. The shape function of a wormhole starts from a finite minimum value, which is select as $b(r_{0})=r_{0}$, where $r_{0}$ represents the wormhole throat radius. The condition ${1-b(r)/r}>0$ ensures that the signature of the wormhole metric remains consistent. To satisfy the flaring-out condition at or near the throat, we should consider $(b(r_{0})-b^{'}(r_{0})r_{0})/b(r_{0})^{2}>0$ \cite{MT metric} or $b^{'}(r_{0})<1$. Since traversable wormholes are examined in this paper which have no horizons therefor the redshift function $\phi(r)$ must be finite everywhere \cite{Sebastian}. For the metric (\ref{wmetric}), the Ricci scalar in the case of non-vanishing redshift function is:
\bea \label{Ricci scalar}
R&=&-\frac{2}{r^{2}} \Big[ (\varphi(r)''r^{2}+2{\varphi(r)'}^{2}r^{2})(1-\frac{b(r)}{r}) 
-\frac{\varphi(r)'}{2}(b(r)'r-b(r))  -{\varphi(r)'}^{2}r^{2}(1-\frac{b(r)}{r}) \nonumber \\ &+&2\varphi(r)'r(1-\frac{b(r)}{r})-
r(\frac{b(r)'}{r}-\frac{b(r)}{r^{2}})-\frac{b(r)}{r} \Big ],
\eea
We consider anisotropic distribution of matter for which the energy-momentum tensor is in the form\cite{Lobo:2009ip}:
\be
T_{\mu\nu}=(\rho+p_t)U_\mu \, U_\nu+p_t\,
g_{\mu\nu}+(p_r-p_t)\,\chi_\mu \chi_\nu \,,
\ee
where $\rho(r)$ is the energy density and  $p_{r}(r)$ and $p_{t}(r)$ are the radial and tangential 
pressures. $U^{\mu}$ is the four-velocity and $\chi^\mu=\sqrt{1-b(r)/r}\,\delta^{\mu}_r$ is the unit spacelike vector in the radial direction. In this paper we investigate the NEC and WEC for the wormhole solutions in the $f(R)$ gravity. These conditions are defined using the formula $T_{\mu \nu}k^{\mu}k^{\nu}\geq0$, where $k^{\mu}$ stands for any null (light-like) vector field for the NEC and any time-like vector field for the WEC  \cite{Hawking}.

Now, by employing Einstein tensor (\ref{Einstein tensor}), we can calculate $f(R)$ gravity field equations for the wormhole geometry as:
\be \label{fe1}
\frac{b'(r)}{r^{2}}f_{R}= \chi(r) f_{RR} 
R'(r)\varphi'(r)+\frac{1}{6}(R(r)f_{R} + f) \\ 
+ \frac{1}{3}(2\rho + 2p_{t} + p_{r}),
\ee

\bea \label{fe2}
\frac{b(r)+ 2r^{2}\varphi'(r)\chi(r)}{r^{3}}f_{R}
&=& (R''(r)\chi(r)+\frac{R'(r)\chi^{'}(r)}{2})f_{RR}+R(r)^{'2}\chi(r)f_{RRR} \nonumber \\ 
&+& \frac{1}{6}(R(r)f_{R}+f)- \frac{1}{3}(\rho + 2p_{r} - 2p_{t}),
\eea

\bea \label{fe3}
\Big(\varphi''(r)+\varphi'(r)^{2} 
+\frac{\varphi'(r)}{r} -\frac{\chi'(r)(r+1)}{2(r-b)}\Big)f_{R}
&=& \frac{R'(r)}{r}f_{RR}+\frac{1}{6\chi(r)}(R(r)f_{R}+f) \nonumber \\ 
&-& \frac{1}{3\chi(r)}(\rho - p_{r} + p_{t}),
\eea
Here $\chi(r)=\frac{b(r)}{r}-1$ and prime denotes derivative with respect to $r$.
The field equations (\ref{fe1})-(\ref{fe3}) can be expressed in form of the energy conditions (EC):

\bea \label{rho}
\rho(r)&=& \chi(r){R'}^{2}(r)f_{RRR} +\Big(\chi(r)R''(r)+(\frac{\chi'(r)}{2}+\frac{2\chi(r)}{r})R'(r)\Big)f_{RR} \nonumber \\
&+&\Big((\frac{\chi'(r)}{2}+\frac{2\chi(r)}{r})\varphi'-\chi(r)(\varphi''(r)+\varphi'^{2}(r))\Big)f_{R}+\frac{f(R)}{2},
\eea

\bea \label{wec1}
wec_{1}(r)&=&\chi(r){R'}^{2}(r)f_{RRR} + \Big((R''(r)-\varphi'(r)R'(r))\chi(r)+\frac{\chi'(r)}{2}R'(r)\Big)f_{RR}\nonumber \\
&+& \Big(\frac{\chi'(r)}{r}-2\frac{\varphi'(r)}{r}\chi(r)\Big)f_{R},
\eea

\bea \label{wec2}
wec_{2}(r)&=& \chi(r)\Big(\frac{1}{r}-\varphi'(r)\Big)R'(r)f_{RR} +\Big((\frac{\chi'(r)}{2}+\frac{\chi(r)+1}{r})(\frac{1}{r}-\varphi'(r)) \nonumber \\
 &-&\chi(r)(\varphi''(r)+\varphi'(r)^2)+\frac{\varphi'(r)}{r}\Big)f_{R} ,
\eea

Where $wec_{1}(r)=\rho(r)+p_{r}(r)$, $wec_{2}(r)=\rho(r)+p_{t}(r)$. At the wormhole throat we obtain:
\bea \label{rho at the throat}
\rho(r)\big|_{r=r_{0}}&=& \frac{R'(r)}{2r}(b'(r)-1)f_{RR}-\frac{\varphi'(r)}{2r}(b'(r)-1)f_{R}+\frac{f(R)}{2},
\eea
\bea \label{wec1 at the throat}
\rho(r)+p_{r}(r)\big|_{r=r_{0}}&=&\frac{R'(r)}{2r}(b'(r)-1)f_{RR}+\frac{1}{r^2}(b'(r)-1)f_{R},
\eea 
\bea \label{wec2 at the throat}
\rho(r)+p_{t}(r)\big|_{r=r_{0}}&=&(\frac{1}{2r}(b'(r)+b(r))(\frac{1}{r}-\varphi'(r))+\frac{\varphi'(r)}{r})f_{R} ,
\eea
It is worth to mention that (\ref{rho at the throat}), (\ref{wec1 at the throat}) and (\ref{wec2 at the throat}) are in agreement with \cite{Pavlovic Sossich 2015}. Existence of $\varphi(r)$ and its derivatives in Equations (\ref{rho at the throat})-(\ref{wec2 at the throat}) means that by selecting the adequate redshift function, as we have done in the following, one may find  wormhole solutions which respect the NEC and WEC around the wormhole throat.  
\section{Traversable wormhole solutions in f(R) models} \label{s3}
In this section, we are interested in studying WEC and NEC near the throat of wormholes that are the solutions for the equations of the motion of four $f(R)$ gravity models:  $\rom{1}$) Exponential gravity model  (\cite{Cognola:2007zu},\cite{Elizalde:2010ts}) , $\rom{2}$) Tsujikawa model \cite{Tsujikawa:2007xu}$\rom{3}$) Starobisky model \cite{star07}, $\rom{4}$) Hu-Sawicki model \cite{HuSa Equation}. 
These models, which are consistent with the cosmological observations, can be written in  the general form \cite{Tsujikawa:2007xu}: 
\be\label{GF f(R)}
f(R)=R-\alpha\xi(R) 
\ee

In this context, the function $\xi(R)$ are presented in the table 1, designed to satisfy two conditions. First, $\xi(R)\big|_{R=0}=0$ our models reduce to the Einstein general relativity. Second, $\xi(R)\big|_{R>>R_{0}}=constant$ to upholding local gravity constraints where $R_{0}$ present background curvature of the universe \cite{Louis Yang}. The parameter  $\alpha$ is a free parameter. In these models, the curvature parameter $R_{*}$ is a small positive constants \cite{Quasi}. Additionally, $\lambda$ and $n$ are free positive parameters  utilized in these models \cite{Tsujikawa:2007xu}.
\begin{table}[h!] 
	
	\begin{center}
		
		\begin{tabular}{|c | c|}
			
			\hline
			
			Model&      $\xi(R)$ \\ \hline
			
			Starobisky&	  $-\lambda R_{*}[1-(1+\frac{R^{2}}{R_{*}^{2}})^{-n}]$\\  [5mm] \hline  
			
			Hu-Sawicki&	 $-\lambda R_{*}\frac{(\frac{R}{R_{*}})^{2n}}{(\frac{R}{R_{*}})^{2n}+1}$\\  [5mm] \hline
			
			The exponential gravity&	 $-\lambda R_{*}(1-e^{-\frac{R}{R_{*}}})$\\ [5mm]  \hline
			Tsujikawa&	 $-\lambda R_{*}tanh(\frac{R}{R_{*}})$\\ [5mm]  \hline
			
		\end{tabular}
		
	\end{center}
	\caption{$f(R)$ gravity models\label{tab:f(R) models}}
	
\end{table}\\
In our investigation, we specifically insert the exponential shape function in conjunction with a non-zero redshift function that is characterized by a fractional formula:
 \be \label{b(r)}
 b(r)=\frac{r}{e^{r-r_{0}}}
\ee 
 \be \label{phi(r)}
\phi(r)=\frac{\varphi_{0}}{r^m}
\ee 
Where $\varphi_0$ is a constant and $r_{0}$ is the throat location. The flare-out condition for the shape function (\ref{b(r)}) is checked by the minimality of the wormhole throat as:
  \be \label{flare-out}
\frac{d}{dz}\Big(\frac{dr}{dz}\Big)=\frac{b-b^{'}r}{2b^2}=\frac{1}{2}>0
 \ee
Inserting the shape function (\ref{b(r)}) and red shift function (\ref{phi(r)}), we can rewrite the Ricci scalar (\ref{Ricci scalar}) in the following way:
 \be \label{Specially Ricci Scalar}
R=-\frac{2m^2\varphi_{0}^2}{r^{2m+2}}\big(1-e^{r_{0}-r}\big)+\frac{2m\varphi_{0}}{r^{m+2}}\big(1-m-e^{r_{0}-r}(1-m-\frac{r}{2})\big)+\frac{e^{r_{0}-r}}{r}\big(\frac{1}{r}-1\big)
\ee
Now, by applying the form of $f(R)$ (\ref{GF f(R)}) in EC equations (\ref{rho})-(\ref{wec2}), we obtain:
\bea \label{wec1 xhi}
wec_{1}(r)&=&\frac{\alpha e^{r_{0}-r}}{2 r^{6}}  \Big[\phi_{0}mr^{-m}e^{r_{0}-r}\big (2r^{-m + 1} + r(3m- 1) + 4r^{-m}(m + 1)+2(m + 2)(m - 1)\nonumber \\
&-&1\Big)- 1 \Big]^{2}\xi_{RRR} +\frac{\alpha e^{-2(r-r_{0})}}{2r^{3m+4}}\Big[4r^{3m}\Big(e^{r-r_{0}}(r^3 + r^2 - 2r - 6)-\frac{r^3}{2} - r^{2} + r + 6\Big)\nonumber \\
&+&  4\phi_{0}mr^{2m}\Big(m^{3} + (2r + 4)m^2 + (\frac{5r^{2}}{4} + 3r + 1)m + \frac{r^3+5r^{2}}{4}-2r-8\Big)\nonumber \\
&-&2\Big(m^3+(r + 4)m^2 + (r^2 + \frac{3}{2}r + 1)m + \frac{r^3+2r^{2}}{4}  - r - 7\Big)e^{r-r_{0}}\nonumber \\
&+& e^{2(r-r_{0})}(m - 1)(m + 3)(m + 2)+12\phi_{0}^{2}m^{2}r^m\Big(m^2 + (\frac{r}{2} + 3)m +\frac{7r}{6} + \frac{8}{3}\nonumber \\
&-&2e^{r-r_{0}}(m^2 + (\frac{r}{4} + 3)m + \frac{r^{2}+7r}{12} +\frac{8}{3})+e^{2(r-r_{0})}(m^2 + 3m + \frac{8}{3})\Big)\nonumber\\
&-&  8\phi_{0}^3m^{3}(e^{r-r_{0}} -1)\Big((m + 1)e^{r-r_{0}}-m - \frac{r}{2} - 1 \Big)\Big]\xi_{RR}\nonumber\\
&-&\frac{e^{r_{0}-r}}{r^{m+2}}\Big(r^{m+1}+2m\phi_{0}(e^{r-r_{0}}-1)\Big)(1-\alpha\xi_{R}),
\eea 

\bea \label{wec2 xhi}
wec_{2}(r)&=&\frac{4\alpha(e^{r_{0}-r}-1)(\phi_{0}m+r^{m})}{e^{r}r^{3m+4}} \Big[r^{2m}e^{r_{0}}(\frac{r^2}{2}-1) +\frac{\phi_{0}mr^{m}}{2}\Big((-m^{2}-(\frac{3r}{2}-1)m\nonumber \\
&+&(m+2)(m-1)e^{r}\Big)+\phi_{0}^{2}m^{2}\Big((m+1)e^{r}-(m+\frac{r}{2}+1)e^{r_{0}}\Big)\Big]\xi_{RR}\nonumber\\
&-&\Big[\frac{\phi_{0}m}{r^{m+2}}\Big((e^{r_{0}-r}-1)( -\frac{\phi_{0}m}{r^m}+m+1)+1\Big)\nonumber\\
&-&\frac{re^{r_{0}-r}}{1+\phi_{0}mr^{-m}}(\frac{-1}{2}+\frac{1}{1+\phi_{0}mr^{-m}})\Big](1-\alpha \xi_{R}),
\eea 
In the limit $\alpha \to 0 $, within the context of Einstein theory, the relation of $wec_{1}(r)$ (\ref{wec1 xhi}) at the throat $r=r_{0}$ reveals the violation of the NEC. Consequently, the WEC is also violated at the throat:
\be \label{wec1 at throat alpha}
\rho+p_{r}\big|_{r=r_{0},\alpha=0}=-\frac{1}{r_{0}}<0,
\ee
Due to the advantages of proper length such as its property of invariant measurement, we substitute the shape function (\ref{b(r)}) in the proper length one finds that:
\be \label{proper length}
l(r)=\pm\int^{r}_{r_{0}}\frac{dr}{\sqrt{1-b(r)/r}}=\pm ln(-\frac{1}{2}+e^{r-r_{0}}+\sqrt{e^{2(r-r_{0})}-e^{r-r_{0}}}) + C,
\ee
Where $C=\pm ln(2)$. One can also find the reverse function $r(l)$  as:
\be \label{r(l)}
r(l)=\pm ln(\frac{1}{4}+\frac{1}{4}(e^{2l}+2e^{l}))+r_{0}-l,
\ee
Note that the positive sign corresponds to the upper universe, while the negative sign corresponds to the lower universe and $l=0$ is equivalent to the throat $r=r_0$. In the following sections, we use the equation (\ref{r(l)}) to plot the energy conditions based on the parameter $l$.
\subsection{Case $\rom{1}$: The Exponential gravity model} \label{Exponential model}
We begin our investigation with the exponential $f(R)$ gravity model, which represented in the following form \cite{Geng}:
\be \label{Exponential f(R)}
\xi(R) =-\lambda R_{*}(1-e^{\frac{R}{R_{*}}})
\ee
Where $\lambda$ is a free positive dimensionless constant and $R_{*}>0$ is a curvature parameter \cite{Quasi}. The exponential $f(R)$ model is a valuable tool for describing the dynamics of galactic phenomena \cite{S. Capozziello 2007}.\\ 
We are interested in wormhole solutions that avoid the anti-gravitating behavior and also without the cosmic instability.To this end we calculate $f_R$ and $f_RR$ at the wormhole throat by utilizing the Ricci scalar \ref{Specially Ricci Scalar}:
\be \label{Ex fR}
f_{R}\big|_{r=r_{0}}=1+\lambda exp\big(\frac{m \varphi_{0}}{R_{*} r_{0}^{m+1}}-\frac{2}{R_{*} r_{0}}(\frac{1}{r_{0}}-1)\big)
\ee
\be \label{Ex fRR}
f_{RR}\big|_{r=r_{0}}=\frac{\lambda}{R_{*}}exp\big(\frac{m \varphi_{0} r_{0}^{1-m}-2r_{0}+2}{R_{*} r_{0}^2}\big)
\ee
Equations (\ref{Ex fR}) and (\ref{Ex fRR}) reveal that regardless of the presence or absence of the redshift function, the  $f_{R}$ and $f_{RR}$ are always positive due to the positive values of the free parameters in the model. So in the exponential $f(R)$ model, there is no ghost solutions and we have cosmological stability. In figure (1.a), $f_{R}$ and $f_{RR}$ are plotted around the throat around the throat when $\varphi_{0}=0$. Note that when  $\varphi_{0}=0$, we explore wormhole geometries without the redshift function. In figure (2.a), $f_{R}$ and $f_{RR}$ are plotted for the case with $\varphi_{0}=1$.\\
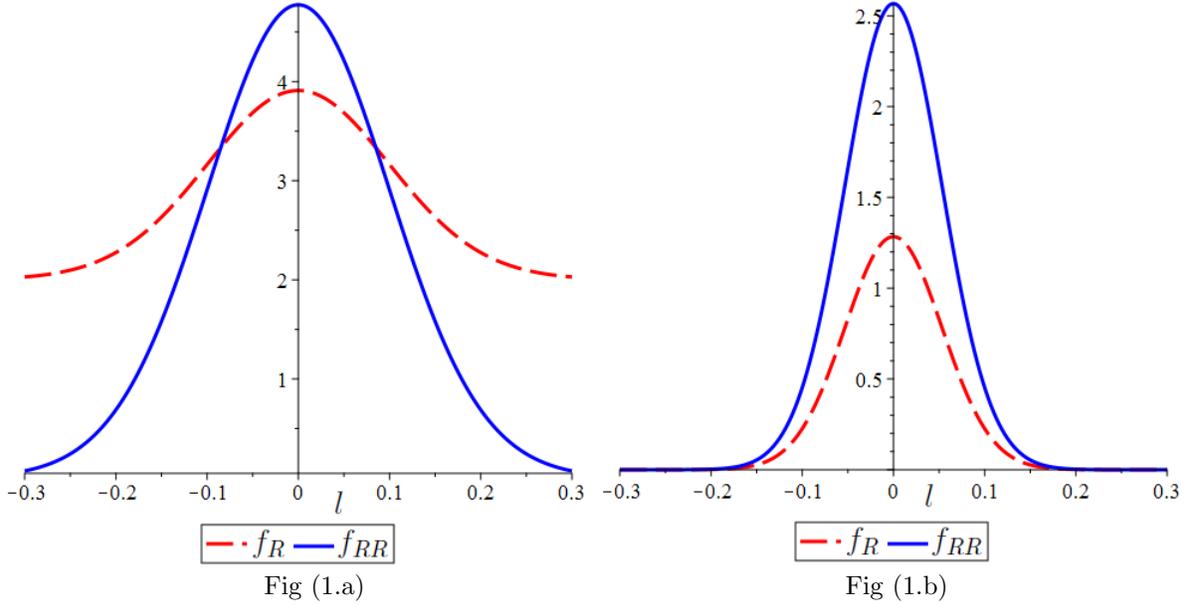
\begin{figure}[H]
	\begin{picture}(0,0)(0,0)
		\put(105,-9){\footnotesize Fig (1.a)}
		\put(325,-9){\footnotesize Fig (1.b)}
	\end{picture}
	\centering
	\includegraphics[scale=0.6]{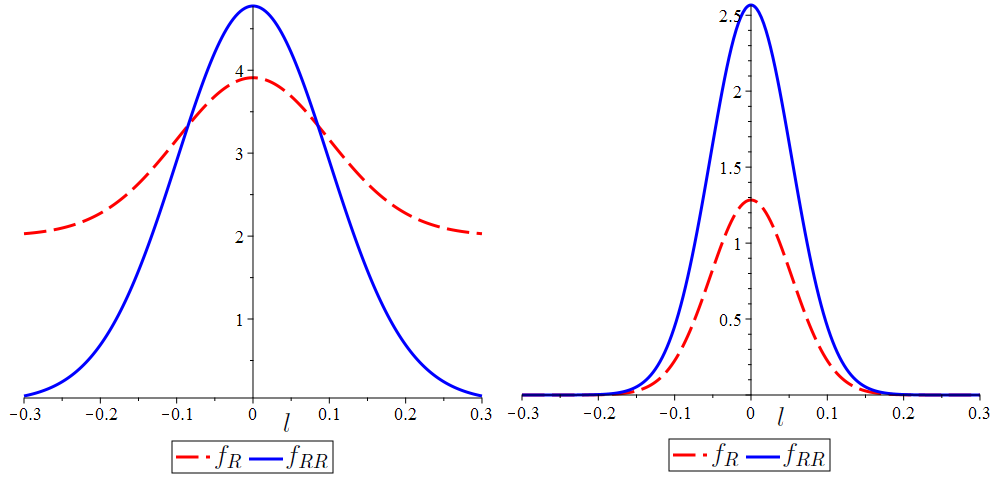}
	\caption{Fig1.a shows that $f_{R}>0$ and $f_{RR}>0$
		for the Exponential model with $\varphi(r)=0$, Fig9.b shows that $f_{R}>0$ and $f_{RR}>0$ when $\varphi_{0}=1$. In these figures we set $\lambda=0.955$, $R_{*}=0.01$, $m=1$ and $r_{0}=1$.}
\end{figure}
In the following, we derive the expressions for energy density $\rho$, $wec_{1}$, and $wec_{2}$ at the wormhole throat by using equation (\ref{wec1 xhi}),(\ref{wec2 xhi}):
\bea \label{Ex rho}
\rho\Big|_{r=r_{0}} &=&\frac{1}{2R_{*}r_0^3}\Big[-r_0R_{*}(R_{*}\lambda r_0^2 + 2r_0 - 2)+\lambda\big(\varphi_{0}m(1-R_{*})r_0^{2 - m}\\ \nonumber
&+& \varphi_{0}(3m^2 - m)r_0^{1-m}  + 2\varphi_{0}^2 m^2r_0^{1-2m} - 2r_0^2 + 4+R_{*}^2r_0^3\big) e^{\eta}\Big]
\eea
\bea \label{Ex wec1}
wec_{1}\Big|_{r=r_{0}} &=&\frac{1}{2R_{*}r_0^3}\Big[\lambda\big(2\varphi_{0}^2m^2r_0^{1-2m}+\varphi_{0}(3m^2 - m)r_0^{1-m}\\ \nonumber
&+&\varphi_{0} m r_0^{2 - m} -2r_0^2(1+R_{*}) + 4\big)e^{\eta}-2r_0^2R_{*}\Big]
\eea
\be \label{Ex wec2}
wec_{2}\Big|_{r=r_{0}} =-\frac{\varphi_{0}mr_0^{1-m} +r_0 -2 }{2r_{0}^2}\big(1+\lambda e^{\eta}\big)
\ee
where $\eta=\frac{\varphi_{0}mr_0^{1-m} - 2r_0 + 2}{r_0^2R_{*}}$. Using equations (\ref{Ex rho}) to (\ref{Ex wec2}), one can find some conditions  to prevent the violation of the NEC and WEC. For instance, equation (\ref{Ex wec2}) indicates that for $r_{0}>2$ in the absence of the redshift function, the  energy conditions will be violated. Ultimately, by choosing appropriate parameters, cases can be found where the energy conditions are maintained. In Figure (2.a), by choosing suitable parameters for a desired case, $f_{R}$ and $f_{RR}$ are plotted that show no anti-gravity solution.  Energy conditions for this case are plotted in figure (2.b) and we can see that NEC and WEC are satisfied.
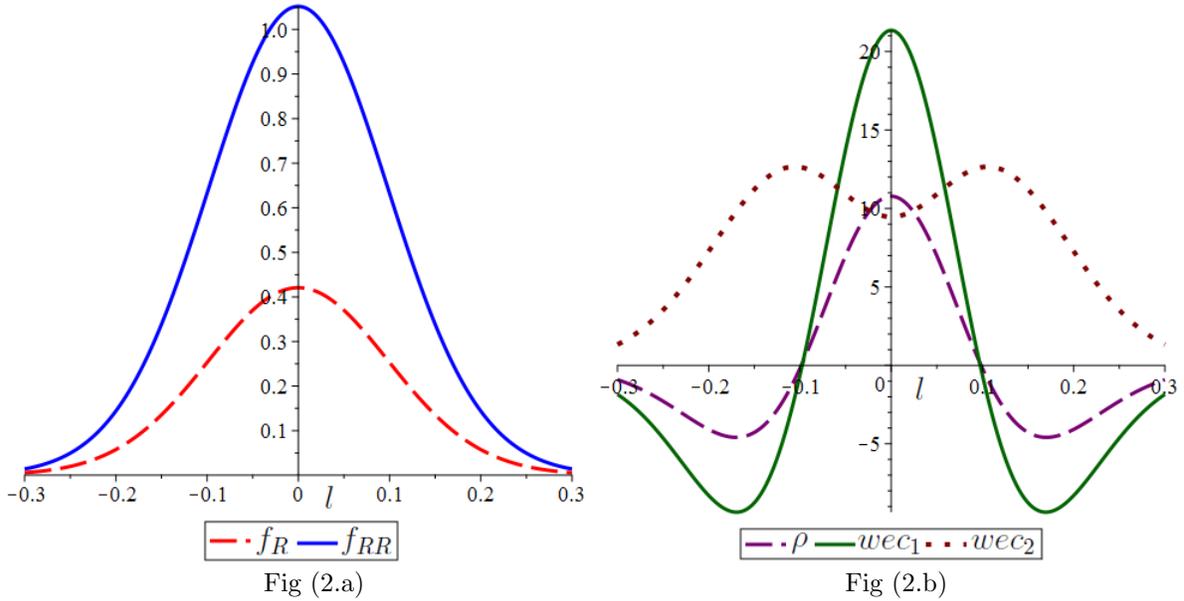
\begin{figure}[H]
	\begin{picture}(0,0)(0,0)
		\put(105,-9){\footnotesize Fig (2.a)}
		\put(325,-9){\footnotesize Fig (2.b)}
	\end{picture}
	\centering
	\includegraphics[scale=0.6]{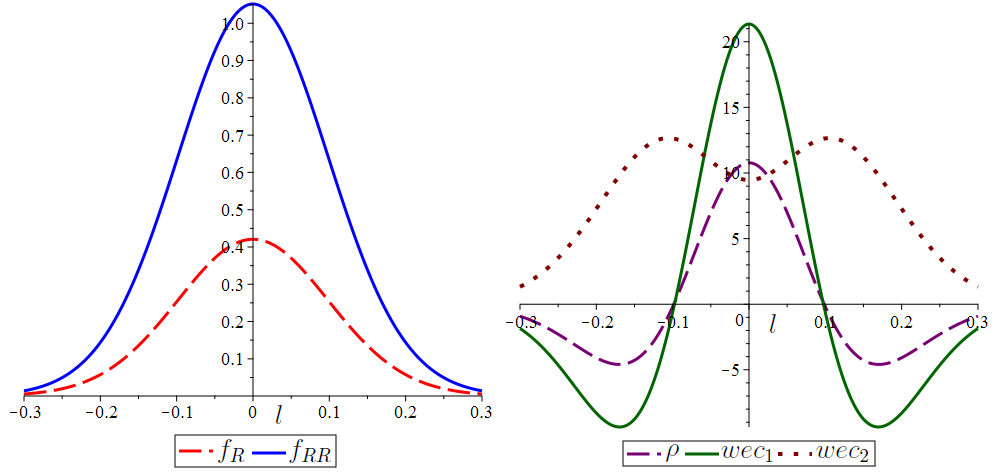}
	\caption{Fig2.a shows that both of $f_{R}$ and $f_{RR}$ are positive,  Fig2.b shows that $\rho(l)>0$, $wec_{1}(l)>0$ and $wec_{2}(l)>0$ in the Exponential model. we set $m=0.1$, $\lambda=0.955$, $R_{*}=0.01$, $r_{0}=1$ and $\varphi_{0}=1$ in these figures. }
\end{figure}
\subsection{Case $\rom{2}$: The tsujikawa model} \label{tsujikawa model}
The second $f(R)$ gravity model that we investigate its wormhole solutions is the Tsujikawa model which is given by:
\be \label{Tsujikawa f(R)}
\xi(R) =-\mu R_{*} tanh(\frac{R}{R_{*}})
\ee
In this model, $\mu$ is dimensionless model parameter where  $0.905<\mu<1$ \cite{Tsujikawa:2007xu}. The Tsujikawa model can be considered as a similar kind of the exponential $f(R)$ model \cite{Ali}, But due to the difference in its functional structure, it leads to different cosmological results \cite{Cosmological evolutions in Tsujikawa model of f(R) Gravity}.\\
Similar to previous models, we start by deriving the expressions for the $f_{R}$ and $f_{RR}$ at the throat to analyze their signs.
\be \label{Tsu fR}
f_{R}\big|_{r=r_{0}}=\frac{ e^{\Omega_1}(e^{\Omega_1}+2-4\mu)+1}{R_{*}(e^{\Omega_2}+1)^2}
\ee
\be \label{Tsu fRR}
f_{RR}\big|_{r=r_{0}}=\frac{8\mu e^{\Omega_1}}{R_{*}(e^{\Omega_2}+1)^3}(e^{\Omega_1}-1)
\ee
Where functions $\Omega_1$ and $\Omega_2$ is:
\be \label{TsuA}
\Omega_1=\frac{2r_0^{-1 - m}m\varphi_0r_{0}^2 - 4(r_0 -1)}{R_{*}r_0^2}\,, \qquad \Omega_2=\frac{2r_0^{1 - m}m\varphi_0 - 4(r_0 -1)}{R_{*}r_0^2}\,.
\ee
Based on equations (\ref{Tsu fR}) and and (\ref{Tsu fRR}) , the positivity of $f_R$ and $f_{RR}$ depends on the throat radii and constant $\varphi_0$ as depicted in figure 3.
\begin{figure}[H]
	\begin{picture}(0,0)(0,0)
		\put(105,-9){\footnotesize Fig (3.a)}
		\put(325,-9){\footnotesize Fig (3.b)}
	\end{picture}
	\centering	
	\includegraphics[scale=0.6]{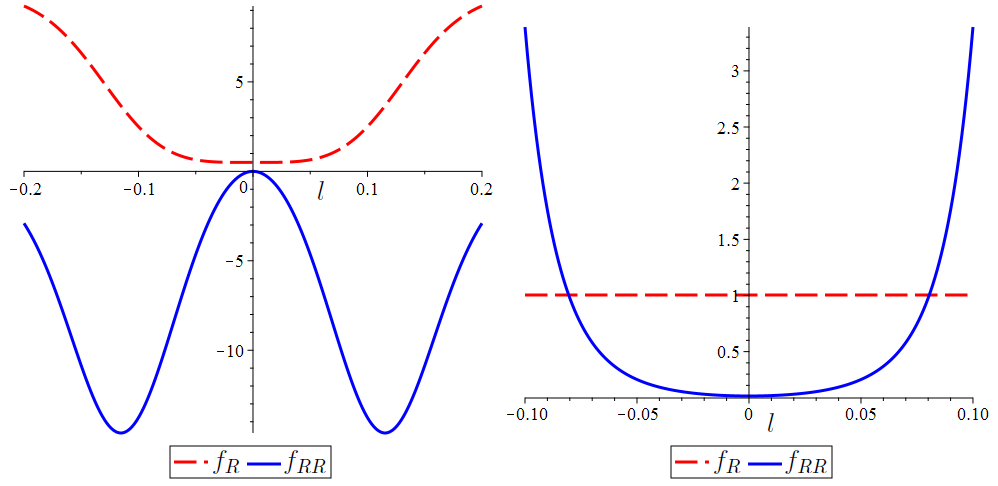}	
	\caption{Fig3.a shows that $f_{R}>0$, $f_{RR}<0$ in Tsujikawa model with $\varphi_0=0$, Fig3.b shows $f_{R}>0$, $f_{RR}>0$ when $\varphi_0=1$, These pictures are plotted by setting $\mu=0.95$, $R_{*}=0.01$ and $r_{0}=1$. }
\end{figure}
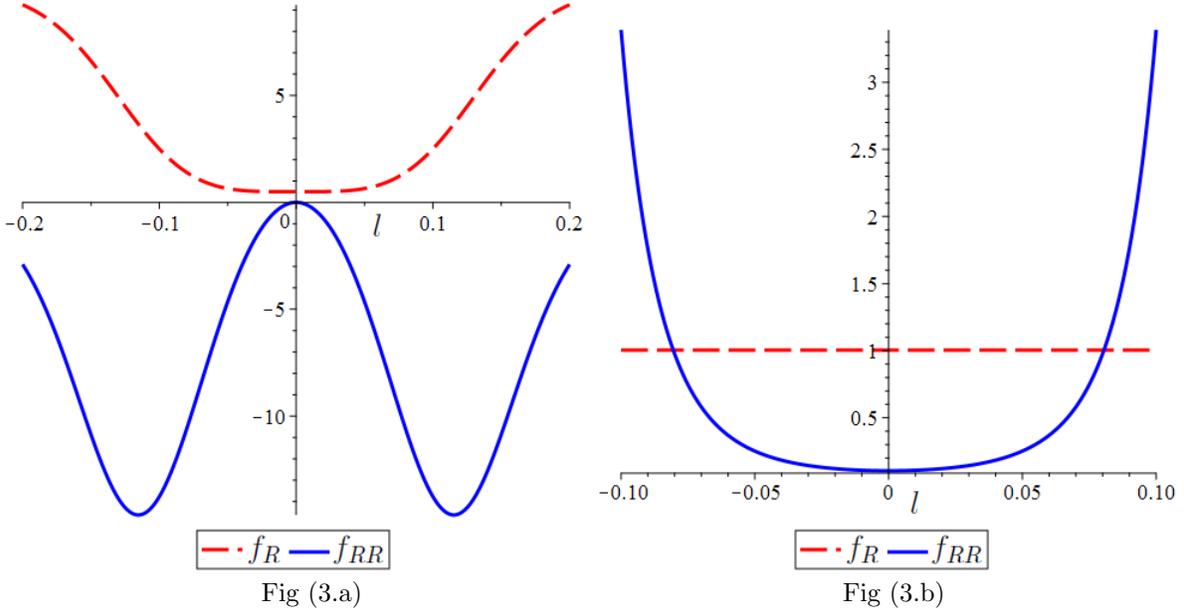
Using equation (\ref{wec1 xhi}), (\ref{wec2 xhi}), we extract the equations for $\rho$, $wec_1$, and $wec_2$ at the throat of wormhole. We then search for  appropriate values that make all these expression positive:
\bea \label{Tsu rho}
\rho\big|_{r=r_{0}}&=&\frac{1}{2R_{*}r_0^3\Omega_3^4}\big[\big(4m\mu\varphi_0(R_{*} - 2)r_0^{2 - m}-\mu\varphi_0(24m^2 - 8m)r_0^{1-m}-8r_0R_{*}(r_0 - 1)\\ \nonumber
&-&16m^2\mu\varphi_0^2r_0^{2m-1} + (2R_{*}^2r_0^3 + 16r_{0}^2 - 32)\mu\big)\Psi(12,2,4)+2\big(2m\mu\varphi_0(R_{*} - 2)r_0^{2 - m} \\ \nonumber
&-&4r_0R_{*}(r_0 - 1)+4\mu\varphi_0(3m^2 - m)r_0^{1-m}+8m^2\mu\varphi_0^2r_0^{2m-1} - (R_{*}^2r_0^3 + 8r_{0}^2-16)\mu\big)\Psi(4,6,12)\\ \nonumber
&+&R_{*}(8mr_0^{2 - m}\mu\varphi_0-12(r_{0}-r_0^2))\Psi(8,4,8)+R_{*}r_{0}(-R_{*}\mu r_0^2 - 2r_0 + 2)\Psi(0,8,16) \\ \nonumber
&+&R_{*}r_{0}(R_{*}\mu r_0^2 - 2r_0 + 2)\Psi(0,0,16)\Big]
\eea
\bea \label{Tsu wec1}
wec_{1}\big|_{r=r_{0}}&=&\frac{1}{R_{*}r_0^3\Omega_3^4}\Big[4\mu\big(\varphi_{0}(m-3m^2)r_0^{1 - m}- \frac{r_0^2R_{*}}{\mu}-2\varphi_{0}^2m^2 r_0^{1 - 2m}\\ \nonumber
&-&  \varphi_{0}mr_0^{2 - m}+(R_{*} + 2)r_0^2-4\big)\Psi(12,2,4)+4\varphi_{0}\mu \big((3m^2-m)r_0^{1 - m}\\ \nonumber
&+&2\varphi_{0}^2m^2 r_0^{1 - 2m}+mr_0^{2 - m}+(R_{*} - 2)r_0^2+4-\frac{r_0^2R_{*}}{\mu}\big)\Psi(4,6,12)\\ \nonumber
&-&R_{*}r_0^2\big((6-8\mu)\Psi(8,4,8)+\Psi(0,0,16)+\Psi(0,8,16)\big)\Big]
\eea
\bea \label{Tsu wec2}
wec_{2}\big|_{r=r_{0}}&=&\frac{2-\varphi_{0}mr_0^{1 - m} - r_0}{2r_0^2 \Omega_3^3}\Big[e^{4r_0+8}(3-4\mu)\big(\Psi(0,4,0)+\Psi(4,2,-4)\big)\\ \nonumber
&+&\Psi(0,0,12)+\Psi(0,6,12)\Big]
\eea
where the function $\Omega_3$ and $\Psi(a_{1},a_{2},a_{3})$ defined as:\\
\be \label{Dr0}
\Omega_3=exp(\frac{2\varphi_{0}mr_0^{1-m}+4}{r_0^2R_{*}})+exp(\frac{4}{r_0R_{*}})
\ee
\be \label{Psi}
\Psi(a_{1},a_{2},a_{3})=\frac{exp(a_{1}r_0 + a_{2}\varphi_0 mr_0^{-1 - m}r_0^2 + a_{3})}{exp(r_0^2R_{*})}
\ee
It is evident from the figure 4 that by selecting a suitable $\varphi_0$, NEC and WEC are respected alongside the conditions for cosmological stability and the absence of ghost solutions.
\begin{figure}[H]
	\begin{picture}(0,0)(0,0)
		\put(105,-9){\footnotesize Fig (4.a)}
		\put(325,-9){\footnotesize Fig (4.b)}
	\end{picture}
	\centering	
	\includegraphics[scale=0.6]{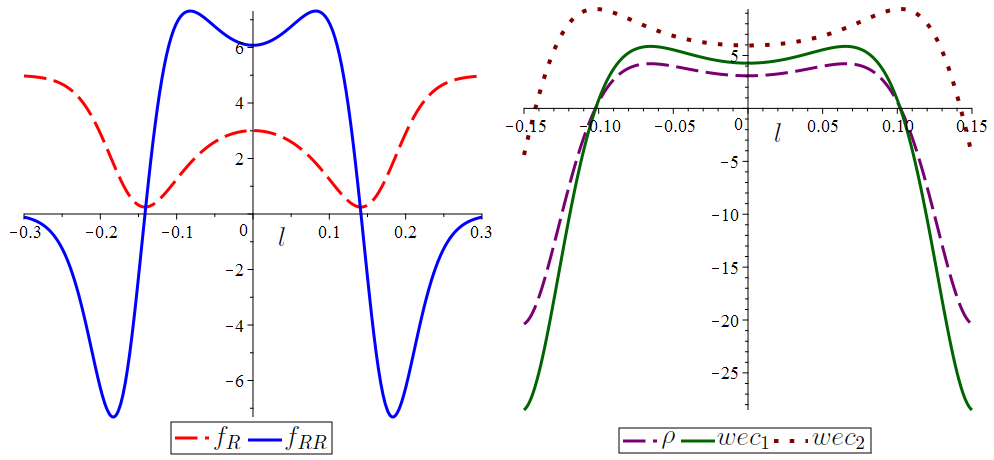}	
	\caption{Fig4.a shows that $f_{R}>0$, $f_{RR}>0$ in Tsujikawa model in the presence of redshift function, Fig4.b shows $\rho(l)>0$, $wec_{1}(l)>0$ and $wec_{2}(l)$, These pictures are plotted by setting $\varphi_{0}=0.01$, $\mu=0.95$, $R_{*}=0.01$ and $r_{0}=1$. }
\end{figure}
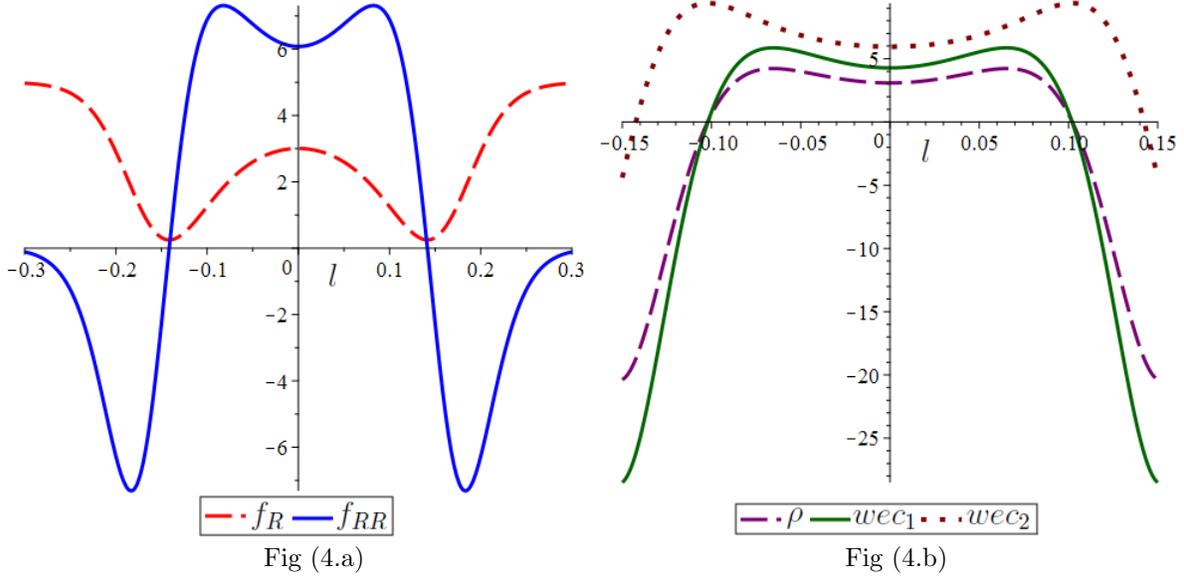

\subsection{Case  $\rom{3}$: The Starobisky gravity model} \label{Starobisky}
Another model that we consider here is the Starobinsky model \cite{star07}. This is a viable modified $f(R)$ gravity model with three free parameters $R_{S}$, $\lambda$ and $n$ \cite{Tsujikawa:2007xu}:
\be \label{Strobinski f(R)}
\xi(R) =-\lambda R_{*}[1-(1+\frac{R^{2}}{R_{s}^{2}})^{-n}]
\ee
Where $\lambda$ is a positive dimensionless parameter which is in the range $0.944<\lambda<0.966$ for $n=2$ \cite{Tsujikawa:2007xu} and $R_{*}$ is a curvature parameter which takes small positive value \cite{Quasi}. The Starobinsky model is one of the important models which describe the cosmic inflation and is consistent with the observations such as the solar system dynamics. Like the two previous models, we are searching for solutions that satisfy NEC and WEC. However, due to the messy form of equations of energy conditions, in this section, we have presented all equations for a specific value of $m$ equal to one. By taking $\lambda$ in the mentioned interval, it is possible to find wormholes that respect NEC and WEC.
Considering the vanishing redshift function, one finds that at the wormhole throat $(l=0)$, $f_{R}$ is positive and $f_{RR}$ is negative which shows that without Redshift function the Starobisky model is unstable despite the absence of ghost solutions:
\be \label{strobinski fR}
f_{R}\big|_{r=r_{0}}= 1+4n\lambda R_{*}^{1-2n}r_0^{2-4n}\sigma_1\big(R_{*}^{2}r_{0}^4+4\sigma_1^2\big)^{-n-1} 
\ee
\bea \label{strobinski fRR}
f_{RR}\big|_{r=r_{0}}&=&\frac{1}{R_{*}^{2}r_{0}^4+4\sigma_1^2}\big[2nr_0^4\lambda(-R_{*}^{2}r_{0}^4+2(2n+1)\sigma_1^2)+R_*(1+\frac{4\sigma_1^2}{R_{*}^{2}r_{0}^4})^{-n}\big]\\ \nonumber
\eea
where $\sigma_1=r_0-1-\frac{\varphi_0}{2}$. In figures (5.a) and (5.b), the  $f_R$ and $f_{RR}$  are plotted in terms of $l$, in the vicinity of wormhole throat. In both cases, the $f_{R}$ is positive, so that we do not have anti-gravity solutions for traversable wormholes.
Note that $f_{RR}$ is negative around the throat in figure (5.a), while by applying the redshift function it becomes positive, resulting in stable cosmological perturbations for our Morris-Thorne wormhole solution as shown in figure (5.b).\\
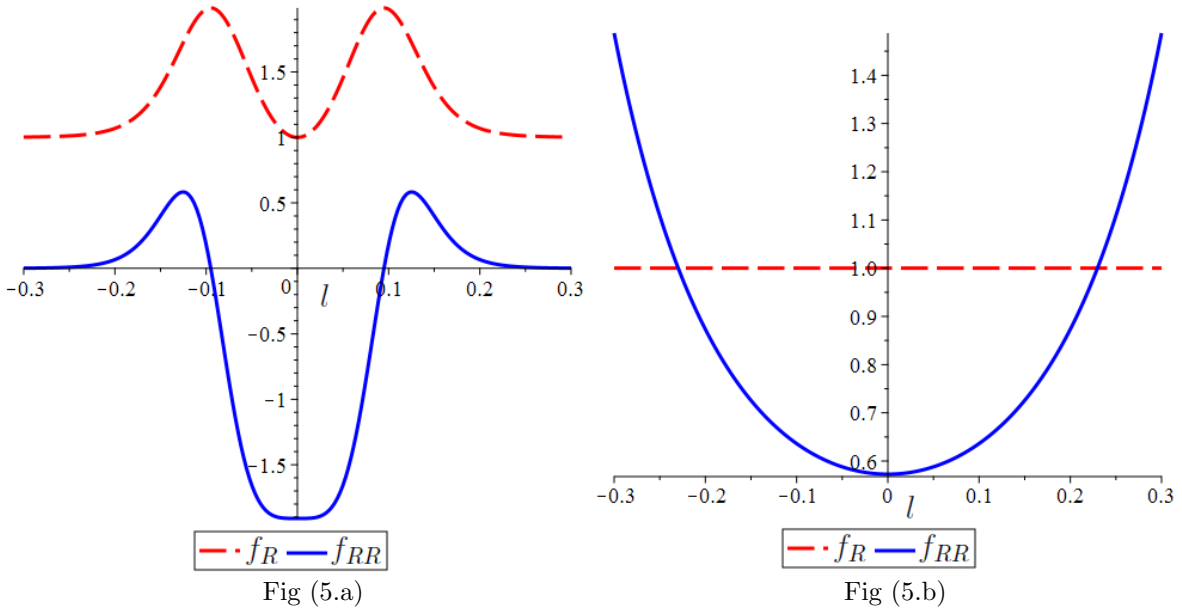
\begin{figure}[H]
	\begin{picture}(0,0)(0,0)
		\put(105,-9){\footnotesize Fig (5.a)}
		\put(325,-9){\footnotesize Fig (5.b)}
	\end{picture}
	\centering
	\includegraphics[scale=0.6]{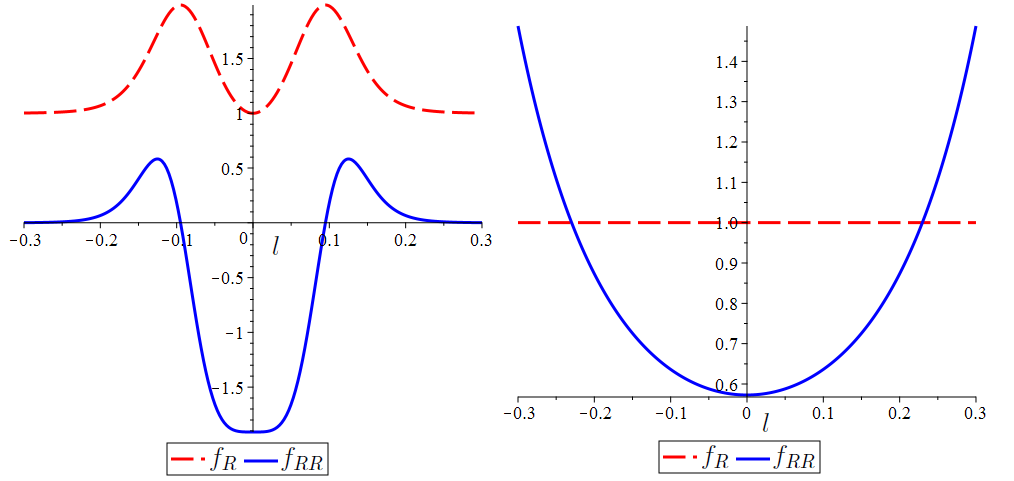}
	\caption{Fig5.a shows that $f_{R}>0$, $f_{RR}<0$ for Starobisky model when $\phi(r)=0$, Fig5.b shows that both of $f_{R}$ and $f_{RR}$ are positive with non-vanishing redshift. we set $R_{*}=0.01$, $n=2$, $\lambda=0.955$, $\varphi_{0}=1$ and $m=1$. }
\end{figure}
By applying equations (\ref{wec1 xhi}) and (\ref{wec2 xhi}), the energy conditions for the wormhole solutions of this model are:\\
\bea \label{Stro rho phi<>0}
\rho\big|_{r=r_{0}}&=&\frac{\sigma_2}{\sigma_3}\big[32\lambda R_{*}r_0^2\sigma_1^2(r_0^3 -\varphi_0^2 - (\frac{r_0^2}{2} + r_0)\varphi_0- 2r_0)n^2+4\lambda R_{*}(\frac{3\varphi_0^4}{2}+\varphi_0^3(\frac{r_0^2}{2}- 6r_0+7)\\ \nonumber
 &+&  (8r_0^2-\frac{r_0^4R_{*}^2}{2} - 3r_0^3- 14r_0 + 10)\varphi_0^2+r_0^2n(4 - \frac{R_*^2r_0^6}{2} - 2r_0^5R_{*}^2 + (R_{*}^2 + 6)r_0^4 - 8r_0^3 - 2r_0^2)\varphi_0\\ \nonumber
 &+& (r_0^3 - 2r_0)(R_{*}r_0^2+2r_0 - 2)(R_{*}r_0^2- 2r_0 + 2)) + \frac{\sigma_3}{2\sigma_2r_0^2}(2(1-r_0) + \lambda R_{*}r_0^2(\sigma_2 - 1))\big], \\ \nonumber
\eea
\bea \label{Stro wec1 phi<>0}
wec_{1}\big|_{r=r_{0}}&=&\frac{1}{2\sigma_3(1-r_0^2)}\big[\frac{\sigma_3}{2r_0^2}-n\lambda R_*r_0\sigma_3\big((-2R_*^2r_0^7 -R_*^2r_0^6(\varphi_{0}+4) + (16n+8)r_0^5  \\ \nonumber
&+& (2R_{*}^2\varphi_0^2- (24n+12)\varphi_0 - 32n)r_0^4 + 12r_0^3(\varphi_0 + 2)(\frac{2n+1}{2}\varphi_0 - \frac{2n+7}{3}) + (8(1-n)\varphi_0^2\\ \nonumber
&-&(2n+1)\varphi_0^3 + (56n+76)\varphi_0 + 32n+80)r_0^2 + 4r_0(\varphi_0 + 2)((3n + 1)\varphi_0^2 -4(n+1)(\varphi_0+1)) \\ \nonumber
&-& 2((2n +1)\varphi_0^2(\varphi_0 + 2)^2))\big)\big],
\eea
\bea \label{Stro wec2 phi<>0}
wec_{2}\big|_{r=r_{0}}&=&\frac{r_0 + \varphi_0-2}{\sigma_3}\big(2\sigma_1^2+2\lambda R_{*}r_0^2n\sigma_1\sigma_2+R_{*}^2r_{0}^4\big),
\eea
where $\sigma_2$ and $\sigma_3$ defined as:
\be \label{Tsu Psi}
\sigma_2=\big(1 +(\frac{2r_0 -\varphi_0-2}{R_{*}r_0^2})^2\big)^{-n}\,, \qquad \sigma_3=2r_0^2\big(r_0^4 R_{*}^2 + (\varphi_0-r_0+1)^2 + 3(r_0-1)^2\big)^2 \,.
\ee
As evident from equations (\ref{Stro rho phi<>0}-\ref{Stro wec2 phi<>0}), in each desired throat radius, with the appropriate determination of the value of $\varphi_0$, related to the redshift function, a stable cosmological wormhole solution can be found that satisfies all three equations and avoids non-gravitational solutions (figure6).

\begin{figure}[H]
	\begin{picture}(0,0)(0,0)
		\put(105,-9){\footnotesize Fig (6.a)}
		\put(325,-9){\footnotesize Fig (6.b)}
	\end{picture}
	\centering
	\includegraphics[scale=0.6]{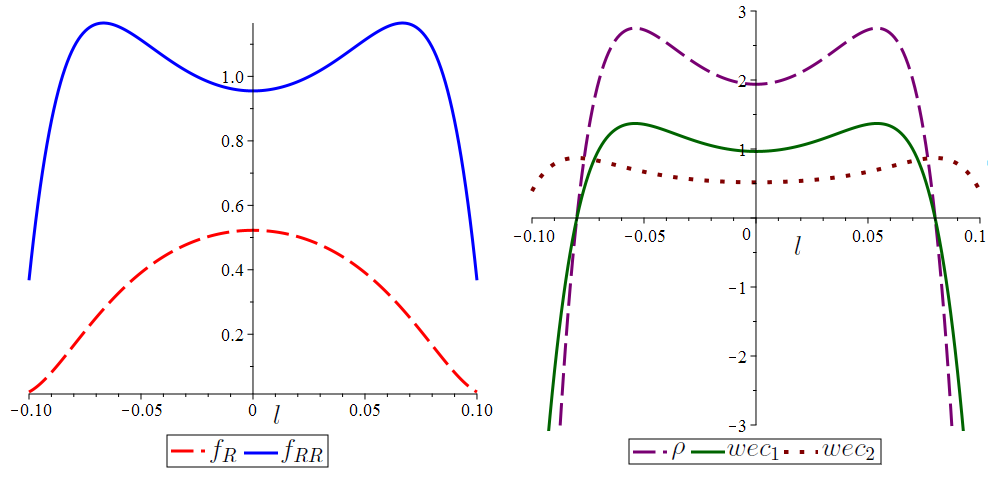}
	\caption{Fig6.a shows that $f_{R}>0$, $f_{RR}>0$. Fig6.b shows that $\rho(l)>0$, $wec1(l)>0$ and $wec2(l)>0$. we set $m=1$, $\varphi_{0}=0.01$, $n=2$, $\lambda=0.955$, $R_{*}=0.01$ and $r_{0}=1$}
\end{figure}
We also explored the outcomes for various values of $m$. Our findings indicate that in the Starobinsky model, the general behavior of $f_R$, $f_{RR}$ ​and the energy conditions remain consistent regardless of the parameter $m$.
\subsection{Case $\rom{4}$: The Hu-Sawicki model} \label{reis}
The last modified gravity model that we consider here, is a cosmologically viable model proposed by Hu and Sawicki \cite{HuSa Equation} which is in the form:
\be \label{Hu-Sawicki f(R)}
\xi(R)=-\lambda R_{*}\frac{(\frac{R}{R_{*}})^{2n}}{(\frac{R}{R_{*}})^{2n}+1}
\ee
Where $n$, $\lambda$ and $R_{*}$ are positive parameters. The Hu-Sawicki model satisfies cosmological and local gravity constraints. In \cite{Tsujikawa:2007xu}, The range of parameters are investigated. It is shown that for $n=1$ one should insert $\lambda\geq8\sqrt{3}/9$. Similar to the previous models, in the first step, we compute $f_R$ and $f_{RR}$ at the wormhole throat but due to the messy form of the energy conditions equation, in the following, we have presented all calculations in the specific case where $m=1$:
\be \label{HuSa fR}
f_{R}\big|_{r=r_{0}}=\frac{(R_*\lambda n r_0^2+2\gamma_1^2)\gamma_2^{2n}+\gamma_1^2(\gamma_2^{4n}+1)}{\gamma_1^2(\gamma_2^{2n}+1)^2}
\ee
\be \label{HuSa fRR}
f_{RR}\big|_{r=r_{0}}=\frac{-2R_*\lambda n r_0^4}{4\gamma_1^2(\gamma_2^{2n}+1)^3}\big(2n(\gamma_2^{2n}-\gamma_2^{4n})-\gamma_2^{2n}-\gamma_2^{4n}\big)
\ee
where $\zeta(r_0)$ and $D(r_0)$ defined as:
\be \label{Tsu ZD}
\gamma_1=r_0 - 1 - \frac{\varphi_0}{2}\,, \qquad \gamma_2=\frac{-2\zeta(r_0)}{r_0^2R_*} \,.
\ee
Considering (\ref{HuSa fR}) and (\ref{HuSa fRR}), one can determine the sign of $f_R$ and $f_{RR}$for different values of the $r_0$. Our research indicates that within suitable radii of the throat which satisfied energy conditions, when the redshift function is absent, the $f_{RR}$ is negative which resulted to cosmic instability. However, by inserting the redshift function, we can find cases where both of $f_R$ and $f_{RR}$ turn positive (Figure 7). 
\begin{figure}[H]
	\begin{picture}(0,0)(0,0)
		\put(105,-9){\footnotesize Fig (7.a)}
		\put(325,-9){\footnotesize Fig (7.b)}
	\end{picture}
	\centering
	\includegraphics[scale=0.6]{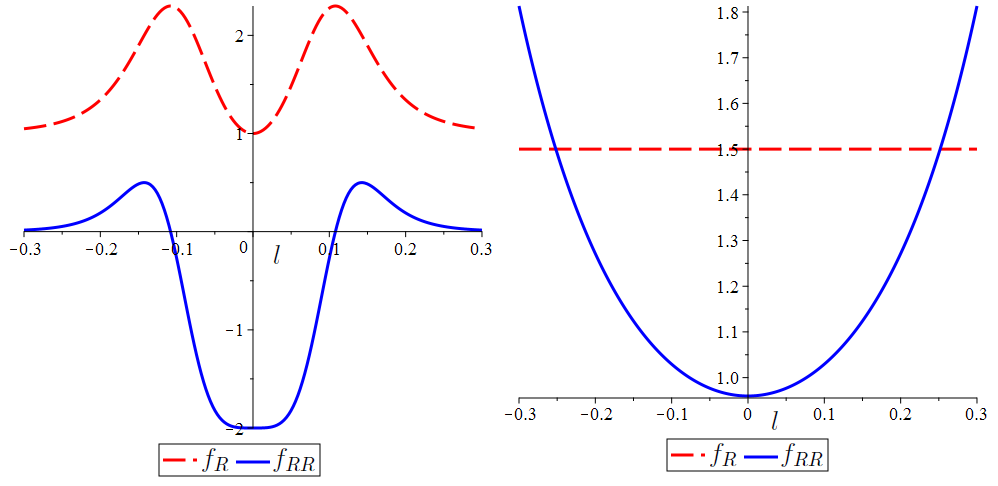}
	\caption{Fig7.a shows that $f_{R}>0$,$f_{RR}<0$ for Hu-Sawicki model when $\phi(r)=0$. Fig7.b shows that $f_{R}>0$, $f_{RR}>0$ when utilizing $\varphi(r)=\frac{1}{r}$. In these cases we set $n=1$, $\lambda=2$, $R_{*}=0.01$ and $r_{0}=1$.}
\end{figure}

Using (\ref{wec1 xhi}) and (\ref{wec2 xhi}), it is straightforward to rewrite $\rho$, $wec_{1}$ and $wec_{2}$ in the following form which, due to their complexity, have been set $m=1$:
\bea \label{Hu rho}
\rho\big|_{r=r_{0}}&=&\frac{1}{2r_{0}^2\gamma_1^2(\gamma_2^{2n}+1)^4}\big[(nr_0^2\lambda R_*(2n\gamma_3-\gamma_4)- (R_*\lambda r_0^2+ 8r_0 - 8)\gamma_1^2)\gamma_2^{2n} \\ \nonumber 
&-&(2r_0^2\gamma_4\lambda R_{*}n + 3\gamma_1^2(R_*\lambda r_0^2 + 4r_0 - 4))\gamma_2^{4n}-(2r_0^2\gamma_3\lambda R_{*}n^2- r_0^2\gamma_4\lambda R_{*}n  \\ \nonumber
&+& 3\gamma_1^2(r_0^2 R_*\lambda + \frac{3}{2}(r_0-1)))\gamma_2^{6n}-4(r_0^2\gamma_2^{8n}R_*\lambda + 2(r_0 - 1)(\gamma_2^{8n} + 1))\gamma_1^2\big]   \\  \nonumber
\eea
\bea \label{Hu wec1}
wec_{1}\big|_{r=r_{0}}&=&\frac{1}{r_{0}\gamma_1^2(\gamma_2^{2n}+1)^4}\big[\big(\frac{n}{2}\lambda r_0R_*(\gamma_3(n-1)-2r_0(r_0-1))\\ \nonumber 
&-& 4\gamma_1^2\big)(\gamma_2^{2n}-\gamma_2^{6n})-(\lambda r_0( \gamma_3+ 2r_0^2 - 2r_0)R_*n - 6\gamma_1^2)\gamma_2^{4n}-2\gamma_1^2(\gamma_2^{8n}+1)\big]\\ \nonumber
\eea
\bea \label{Hu wec2}
wec_{2}\big|_{r=r_{0}}&=&\frac{2-r_0 -\varphi_0}{2r_{0}^2\gamma_1^2(\gamma_2^{2n}+1)^3}\big[(\lambda r_0^2 R_*n + 3\gamma_1)\big(\gamma_2^{2n}+\gamma_2^{4n}\big)+\gamma_1^2(\gamma_2^{6n} + 1)\big]\\ \nonumber
\eea
where $\gamma_3=-\varphi_0^2 - (\frac{1}{2}r_0^2 + r_0)\varphi_0 + r_0^3 - 2r_0$ and $\gamma_4=(r_0^3 -\frac{1}{2}\varphi_0r_0^2 - \frac{3}{2}\varphi_0^2 - 2r_0 - \varphi_0)$. By setting suitable parameters and considering appropriate $\varphi_{0}$ in each throat radius $r_0$, one can find geometries where the NEC and WEC are satisfied.
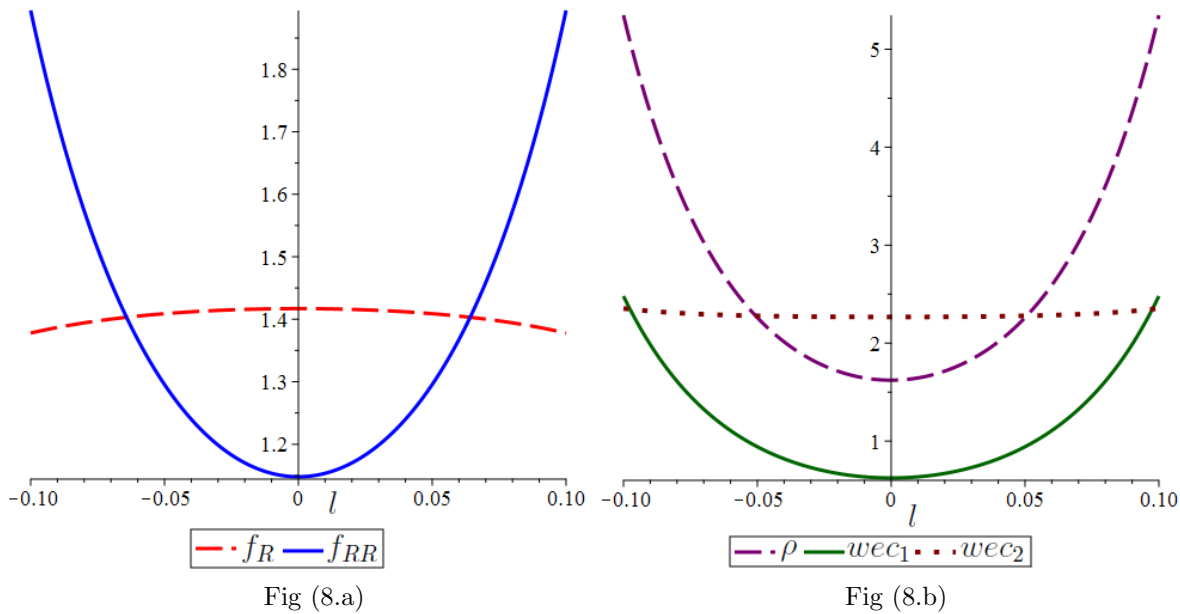
\begin{figure}[H]
	\begin{picture}(0,0)(0,0)
		\put(105,-9){\footnotesize Fig (8.a)}
		\put(325,-9){\footnotesize Fig (8.b)}
	\end{picture}
	\centering
	\includegraphics[scale=0.6]{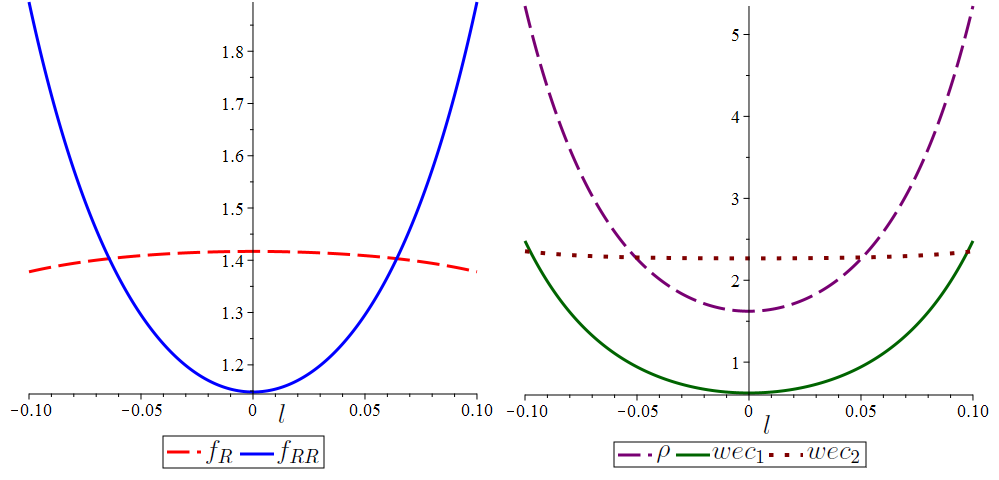}
	\caption{Fig8.a shows that $f_{R}<0$ and $f_{RR}>0$ results in anty-gravity solutions Fig8.b shows that $\rho(l)>0$,$wec_{1}(l)>0$ and  $wec_{2}(l)>0$ at the throat and its vicinity. In these figures, we set $\varphi_{0}=0.04$, $n=1$, $\lambda=2$, $R_{*}=0.01$ and $r_{0}=1$. }
\end{figure}
We also examined the results for different values of $m$. We found that in the Hu-Sawicki model, the generic behavior of $f_R$, $f_{RR}$ and energy conditions are independent of the parameter $m$.
\section{Gravitational lensing and Particle Trajectories Around the Wormhole} \label{s4}
The curved space-time around a wormhole can deflect the light rays and affects the particle trajectories. In this section, by using the effective potential, we find the path of particles around the wormholes. We also analyze the gravitational lensing in the case of f(R) wormholes.

\subsection{Effective Potential}
First, using the Lagrangian formalism, we calculate the effective potential by using the particle trajectories around the wormhole. Taking spherical symmetry and considering the equatorial plane $\theta=\frac{\pi}{2}$, Lagrangian for the metric (\ref{wmetric}) is written as: \cite{lagf}: 
\be \label{lagrangian}
\mathfrak{ L} =g_{\mu\nu} \dot{x}^\mu \dot{x}^\nu=-e^{2\phi(r)}\dot{t}^{2}+\frac{\dot{r}^{2}}{1-\frac{b(r)}{r}}+r^{2}\dot{\phi}
\ee
Where dot refers to the first-order derivative with respect to the affine parameter $\eta$. The Lagrangian (\ref{lagrangian}) on a geodesic is constant which we demonstrate it by $\mathfrak{ L}(x^{\mu},\dot{x}^\mu)=\epsilon$. When $\epsilon=-1$ and $\epsilon=0$ a time-like and null geodesics exits, respectively. Using the Euler-Lagrange equation:
\be \label{Euler-Lagrange}
\frac{d}{d\eta} \frac{\partial{\frak
		L}}{\partial\dot{x}^{\mu}}-\frac{\partial{\frak L}}{\partial
	x^{\mu}}=0,
\ee
For a test particle with energy E and angular momentum L one can derive the following constants of motion:
\bea \label{Motion Constant}
-e^{2\phi(r)}\dot{t}=E, \qquad 2r^{2}\dot{\phi}=L.
\eea
By replacing the motion constants ($E$ and $L$) into the equation (\ref{lagrangian}), we obtain:
\bea \label{Motion Constant r}
\dot{r}^2=e^{-2\phi(r)}(1-\frac{b(r)}{r})(E^{2}-\frac{L^{2}}{r^{2}}).
\eea
Using the proper radial distance(\ref{proper length}), constants of motion(\ref{Motion Constant}) can be written as:
\be \label{system MC}
\dot{l}^2+V_{eff}(L,l)=E^{2},
\ee
Where $V_{eff}$ is the effective potential which determined:
\be \label{Veff}
V_{eff}(L,l)=e^{2\phi(l)}(\frac{l^{2}}{r(l)^{2}}-\epsilon),
\ee
By comparing the effective Potential and total energy $E$ of the particle, the possibility of the particle passing through the wormhole throat can be investigated. If $E^{2}>V_{eff}(L,0)$, the particle passes through the throat and enters another world, otherwise it returns to the primary world when $E^{2}<V_{eff}(L,0)$. In such scenarios, a significant point to consider is the existence of a turning point, denoted as $l=l_{o}$, which can be determined by solving the following equation:
\be \label{orbit}
E^2=V_{eff}(L,l_{o}).
\ee

$V_{eff}(L,l)$ is plotted against $l$ for null (figure 9.a)  and timelike (figure 9.b) geodesics. We investigated the results for different $L$ when $\varphi_{0}>0$. As illustrated in figure 9, Increasing the value of $L$ results in escalation of $V_{eff}(L,0)$.
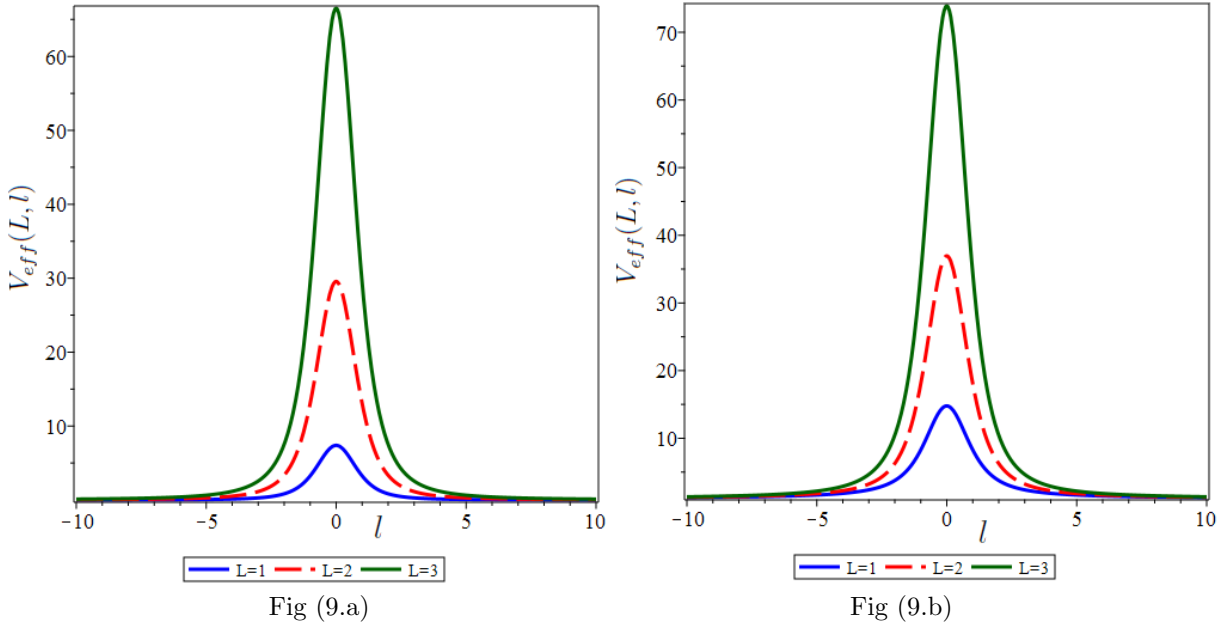
\begin{figure}[H]
	\begin{picture}(0,0)(0,0)
		\put(105,-9){\footnotesize Fig (9.a)}
		\put(325,-9){\footnotesize Fig (9.b)}
	\end{picture}
	\centering	
	\includegraphics[scale=0.6]{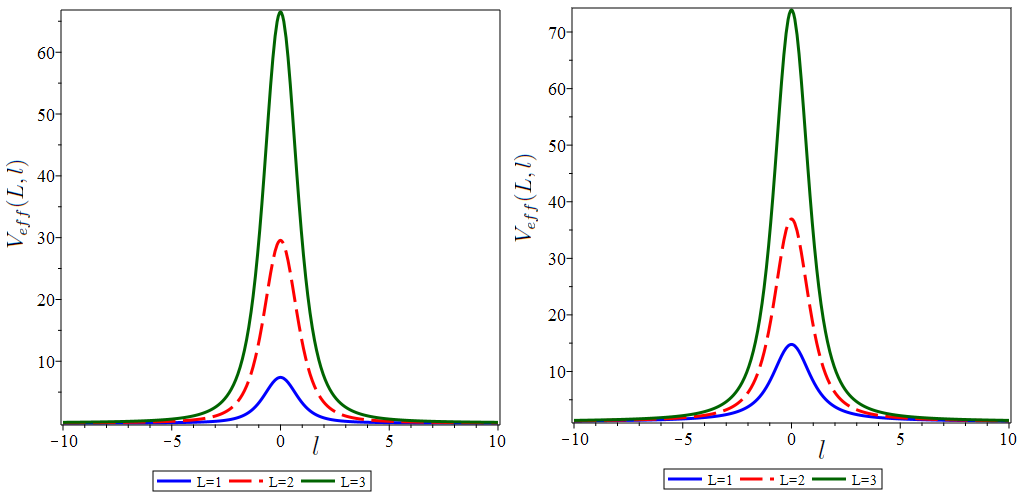}	
	\caption{Fig9.a shows $V_{eff}(L,l)$ in null geodesics, Fig9.b shows $V_{eff}(L,l)$ for timelike geodesics. these curves plot by setting $\varphi_{0}=1$, $r_{0}=1$ and $L$ has three values= 1, 2, 3.}
\end{figure}
As we can see from Figure 9, the pattern of effective potential is the same for both timelike or null geodesics, note that the maximum of $V_{eff}(L,l)$ increases with the growth of L
This means that if the $E^2=V_{eff}(L,0)$ then the particle will move in an unstable circular orbit at the location of wormhole throat.

$V_{eff}(L,l)$ is plotted against $l$ when  $\varphi_{0}<0$  for null (figure 10.a) and timelike (figure 10.b) geodesics. By comparing figures 13.a and 13.b, it is evident that the $V_{eff}(L,l)$ in null geodesics shows different behavior form timelike geodesics. In null geodesics with increasing $L$ the shape of $𝑉_{𝑒𝑓𝑓}$ does not change, but in timelike geodesics with increasing $L$ the $V_{eff}(L,l)$ reaches a relative minimum that creates a state which cause particles to be trapped in the throat of wormhole and revolve on a stable circular orbit.
\begin{figure}[H]
		\begin{picture}(0,0)(0,0)
		\put(105,-9){\footnotesize Fig (10.a)}
		\put(325,-9){\footnotesize Fig (10.b)}
	\end{picture}
	\centering	
	\includegraphics[scale=0.6]{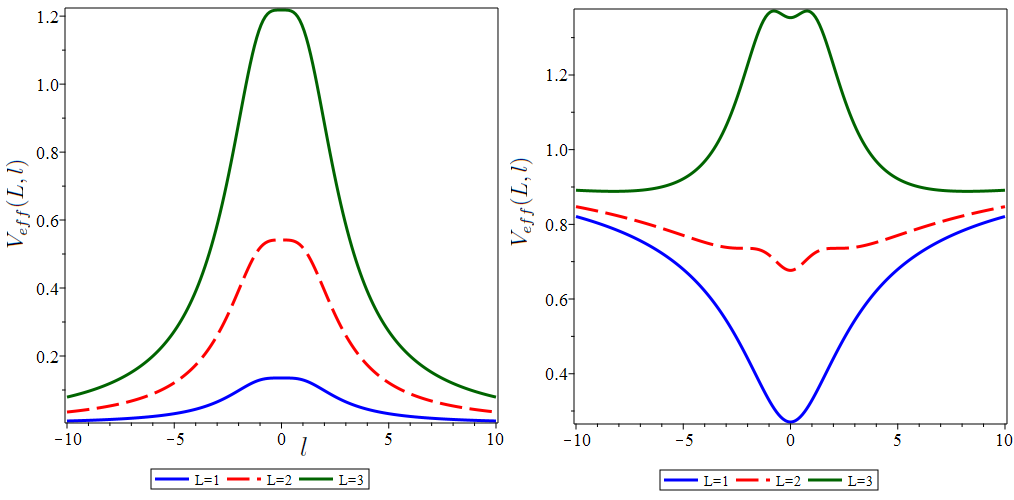}	
	\caption{Fig10.a shows $V_{eff}(L,l)$ in null geodesics, Fig10.b shows $V_{eff}(L,l)$ for timelike geodesics. these curves plot by setting $\varphi_{0}=-1$, $r_{0}=1$ and $L$ has three values:1, 2, 3.}
\end{figure}

\subsection{The Deflection Angle}
When a beam of light passes from infinity to near a massive object like a blackhole, it bends from its direct route and approaches the center of gravitational body to distance $r_{c}$. A wormhole has a strong gravitational 
field, so it acts like a gravitational lens and diverts beam of light. The 
closest path of light near the throat is $r_{c}$. As gravitational field be stronger, the deflection of light beam be greater. The Deflection Angle $\Theta(r_{c})$
is a good 
observational measurement of light bending and obtain from following 
relation for a Morris-Thorne Wormhole metric (\ref{wmetric}) \cite{deflection angle}:

\be \label{Deflection angle}
\Theta_{r_{c}} =-\pi+2\int_{r_{c}}^\infty \frac{e^{\phi(r)}dr}{r^{2}\sqrt{(1-\frac{b(r)}{r})}(\frac{1}{\beta^{2}}-\frac{e^{2\phi(r)}}{r^{2}})},
\ee
Where $\beta$ is Imapct factor and and we have $\beta=r_{c}e^{-\phi(r_{c})}$ for null geodesics. By replacing $\beta$, the shape function (\ref{b(r)}) and the redshift function (\ref{phi(r)}) in  deflection angle (\ref{Deflection angle}), we obtain:

\be \label{Special Deflection angle}
\Theta_{r_{c}}(\phi_{0},m) =-\pi+2\int_{r_{c}}^\infty \frac{e^{\phi_{0}/r^{m}}dr}{r^{2}\sqrt{(1-\frac{1}{e^{r-r_{0}}})}(\frac{e^{2\phi_{0}/r_{c}^{m}}}{r_{c}^{2}}-\frac{e^{e^{2\phi_{0}/r^{m}}}}{r^{2}})},
\ee

\begin{figure}[!h]
	\begin{picture}(0,0)(0,0)
	\end{picture}
	\centering
	\includegraphics[scale=0.6]{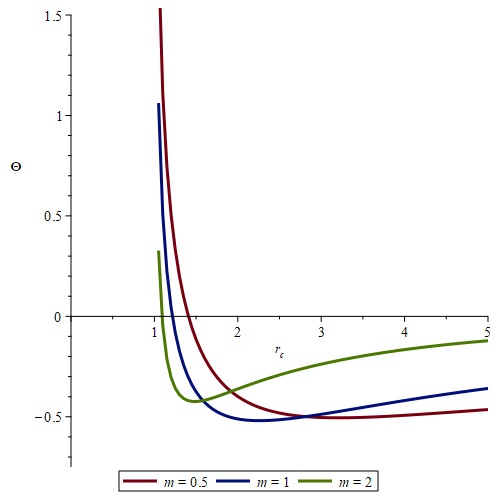}
	\caption{shows the deflection angle for 3 values of parameter $m$, we set $\varphi_0=1$ and $r_0=1$.}
\end{figure}
In the following, we plot the deflection angle $\Theta_{r_{c}}$ according to $r_{c}$ for three different values of $m$. Figure 11 shows that in $r_{0}<r_{c}<\infty$, the deflection angle has finite quantity, and as the distance $r_{c}$ decreases to the wormhole throat, the deflection angle increases. In other words, as the light ray closes to the wormhole throat, where the gravitational field is stronger, its deflection from the original path will be more pronounced. In the wormhole throat, where the gravitational field is extremely strong, the deflection angle tends to infinity. When $r_{c}$ increases to infinity, the deflection angle tends to zero. Figure 11 also illustrates when $r_{c} \to \infty $, for larger values of $m$ the angle of deflection approaches to zero sooner. as shown in figure 11, the value of the parameter $m$ increases, the deflection angle of the light ray near the throat decreases.  Conversely, for smaller values of $m$, the deflection angle near the wormhole throat becomes larger.
\section{Conclusions} 
In this article, we investigate wormhole solutions with non-vanishing redshift function in the framework of $f(R)$ gravity. First, we solved the field equations and then by inserting the exponential shape function and a fractional redshift function we analyzed the wormhole solutions by considering different cosmological viable $f(R)$ models. We rewrote the radial coordinate $r$ in terms of the proper length $l$ and investigated NEC and WEC at the vicinity of wormhole throat. we also took into account the positivity of $f_R$ and $f_{RR}$ to avoid gravitational ghosts and cosmological instability in the solutions. By choosing suitable parameters of each model and redshift function, we looked for the wormhole solutions without the need to exotic matter. We also examined the effective gravitational potential for null and time-like geodesics. Finally, we calculated the deflection angle near the wormhole structures to analyze gravitational lensing. In the following we review the results for these $f(R)$ models:\\
\hspace*{4mm}
case $\rom{1}$: The Exponential model\\
\hspace*{4mm} Considering the applied redshift function, both the $f_R$ and $f_{RR}$ will always be positive. Therefore, the exponential gravity model, does not contain ghost solutions and also our solutions remain cosmologically stable. By selecting appropriate parameters $\varphi_{0}$ and $m$, we found the wormhole solutions that satisfy NEC and WEC, here the parameters $\lambda$ and $R_{*}$ have no considerable influence on the behavior of energy conditions.\\
\hspace*{4mm} case $\rom{2}$: The Tsujikawa model\\
In the Tsujikawa model, contrary to the previous model, the $f_R$ and $f_{RR}$ are not always positive. Particularly, $f_{RR}$ is negative in many cases. However, by inserting the redshift function with suitable values of $\varphi_{0}$ and $m$ , it is possible to find wormhole geometries with cosmological stability. The only free parameter in the Tsujikawa model, $\mu$, has no effect on the satisfaction of the energy conditions and it is easy to find solutions that satisfy NEC and WEC alongside the positivity of $f_R$ and $f_{RR}$ by introducing the redshift function and setting the adequate parameters for it.\\
\hspace*{4mm}
case $\rom{3}$: The Starobinsky model\\
The positivity of $f_R$ and $f_{RR}$ and the satisfaction of NEC and WEC in the Starobinsky model depend on the selection of appropriate parameters and inserting the redshift function facilitates this process. The free parameters of the model $n$ and $\lambda$, do not significantly influence the attainment of the desired solutions. In fact we found the the best way to construct wormhole solutions which respect the NEC and WEC is choosing adequate values for $\varphi_{0}$.\\
\hspace*{4mm}
case $\rom{4}$: The Hu-Sawiciky model\\
The behavior of the Hu-Sawicki model resembles that of the Starobinsky model. By selecting appropriate $\varphi_{0}$
in the redshift function, we were able to find wormhole geometries where the NEC and WEC are satisfied, and the $f_R$ and $f_{RR}$ become positive.
\hspace*{4mm} 

As a final remark, our results showed that incorporating the redshift function into wormhole solutions has a substantial impact on their behavior. Certainly, Introducing the redshift function $\varphi(r)$ prevents the $f_{RR}$ from becoming negative in the tsujikawa, Strabinsky and Hu-sawicky models. Furthermore, by examining the value of the parameters $\varphi_0$ in the $\varphi(r)$, we can find wormhole solutions that exist without the need for exotic matter.
These solutions are stable against spherically symmetric perturbations and provides an interesting possibility to construct non-static or thin-shell wormhole solutions\cite{Bronnikov:2006pt} in the considered f(R) gravity models.

\section*{Acknowledgment}


\begin{thebibliography}{99}
\bibitem{flamm} L. Flamm, Phys. Z. 17, 448 (1916). 
\bibitem{EinstenRosen} A. Einstein and N. Rosen, Phys. Rev. 48, 73 (1935).
\bibitem{misner-wheeler} C. W. Misner and J. A. Wheeler, Ann. Phys. {\bf 2}, 525 (1957); \\C. W. Misner, Phys. Rev. {\bf 118}, 1110 (1959).
\bibitem{MT metric} M. S. Morris and K. S. Thorne, Am. J. Phys. 56, 395 (1988).\\M. S. Morris, K. S. Thorne and U. Yurtsever, Phys. Rev. Lett. 61, 1446 (1988).
\bibitem{ExM Re}S. Kar, N. Dadhich, and M. Visser, Pramana J. Phys. 63, 859 (2004);\\
D. Hochberg and M. Visser, Phys. Rev. D 56, 4745 (1997).
\bibitem{Ex evidence}
S. M. Carroll, Living Rev. Rel. 4, 1 (2001);
\\P. J. E. Peebles and B. Ratra, Rev. Mod. Phys. 75, 559 (2002);
\\V. Sahni, Class. Quantum Grav. 19, 3435 (2002);
\\T. Padmanabhan, Phys. Rep. 380, 235 (2003);
\\P. F. Gonzales-Diaz, Phys. Rev. D 65, 104035 (2002).
\bibitem{Visser}
M. Visser, Nucl. Phys. B328, 203-212 (1989);\\ M. Visser,
Phys. Rev. D39, 3182-3184 (1989);\\ N. M. Garcia,
F. S. N. Lobo and M. Visser, Phys. Rev. D86, 044026
(2012).
\bibitem{modi grav}
T. Clifton, P. G. Ferreira , A. Padilla and C. Skordis, Modified gravity and cosmology, Physics Reports 513 1–189 (2012).
\bibitem{thin shell}
S. H. Mazharimousavi, M. Halilsoy, and Z. Amirabi, Phys.
Rev. D 81, 104002 (2010);\\ Classical Quantum Gravity 28,
025004 (2011);\\ M. R. Mehdizadeh, M. K. Zangeneh, F. S. N. Lobo, Phys. Rev. D 92, 044022 (2015).
\bibitem{harko2013}
T. Harko, F. S. N. Lobo, M. K. Mak and S. V. Sushkov, Modified-gravity wormholes without exotic matter, Phys. Rev. D 87, 067504,  1-5, (2013).
\bibitem{Richarte}
M. G. Richarte and C. Simeone, Thin-shell wormholes supported by ordinary matter in Einstein-Gauss-Bonnet gravity, Phys. Rev. D 76, 087502 (2007).
\bibitem{Alexeyev 2011}
S. O. Alexeyeva, K. A. Rannua, and D. V. Gareevab, Possible Observational Manifestations of Wormholes in the Brans–Dicke Theory, Journal of Experimental and Theoretical Physics,Vol. 113, No. 4, pp. 628–636, (2011).

\bibitem{Horvat}
A. DeBenedictis and D. Horvat, On wormhole throats in f(R) gravity theory, General Relativity and Gravitation volume 44, pages2711–2744 (2012).

\bibitem{Mehdizadeh Lovelock 2016}
M. R. Mehdizadeh and F. S. N. Lobo,Novel third-order Lovelock wormhole solutions,Phys. Rev. D, 93, 124014 – Published 7 June (2016).

\bibitem{Kar2016}
R. Shaikh and S. Kar, Wormholes, the weak energy condition, and scalar-tensor gravity, Phys. Rev. D, 94, 024011 (2016).

\bibitem{Mehdizadeh and Ziaie Einstein-Cartan}
M. R. Mehdizadeh and A. H. Ziaie, Charged wormhole solutions in Einstein-Cartan gravity, Phys. Rev. D 99, 064033 – Published 22 March (2019).

\bibitem{Mehdizadeh and Ziaie cubic}
M. R. Mehdizadeh and A. H. Ziaie, Charged wormhole solutions in Einstein-Cartan gravity, Phys. Rev. D 99, 064033 – Published 22 March (2019).

\bibitem{buchadahl} Buchadahl, H.~A., Mon. Not. Roy. Astron. Soc. 150, 1 (1970).

\bibitem{Pavlovic Sossich 2015} P. Pavlovic, M. Sossich, Wormholes in viable f(R) modified theories of gravity and weak energy condition,The European Physical Journal C, \textbf{75}, (2015).  
\bibitem{Capozziello}
S. Capozziello, M. D. Laurentis, S. D. Odintsov and A. Stabile, Hydrostatic equilibrium and stellar structure in $f(R)$ gravity, Phys. Rev. D, 83, 064004 (2011).
\bibitem{Bakirova}
E. Bakirova and V. Folomeev, Dipole magnetic field of neutron stars in f(R) gravity, General Relativity and Gravitation, 48, (2016).
\bibitem{Napolitano}
N. R. Napolitano, S. Capozziello, A. J. Romanowsky, M. Capaccioli and C. Tortora, Testing Yukawa-like potentials from $f(R)$-gravity in elliptical galaxies,The Astrophysical Journal, 748, (2012).
\bibitem{Terukina}
A. Terukina, L. Lombriser, K. Yamamoto, D. Bacon, K. Koyama, R. C. Nichol, Testing chameleon gravity with the Coma cluster, 	Cosmology and Nongalactic Astrophysics, 2014, (2014).
\bibitem{Cosmological Evolution in f(R)}
J. Q. Guo, A. V. Frolov, Cosmological dynamics in f(R) gravity, Cosmology and Nongalactic Astrophysics, Phys. Rev. D, 88, (2013).	
\bibitem{Planck 2015 results}
Planck Collaboration. Planck Results. XIV. Dark energy and modified gravity. Astronomy and Astrophysics, 594, (2015).
\bibitem{Arnold2015}
C. Arnold, E. Puchwein and V. Springel, The Lyman α forest in f(R) modified gravity, Monthly Notices of the Royal Astronomical Society, 448, (2015).
\bibitem{Cosmological evolutions in Tsujikawa model of f(R) Gravity}
J. Y. Cen, S. Y. Chien, C. Q. Geng and  C. C. Lee, Cosmological evolutions in Tsujikawa model of f(R) Gravity, Physics of the Dark Universe, 26, (2019).
\bibitem{Wh geometries in f(R) modified theories of gravity}
F. S. N. Lobo and M. A. Oliveira, Wormhole geometries in $f(R)$ modified theories of gravity, Phys. Rev. D, 80, (2009).
\bibitem{Eiroa 2016}
E. F. Eiroa, G. F. Aguirre, Thin-shell wormholes with charge in F(R) gravity, The European Physical Journal C, 76, (2016).
\bibitem{Quasi} 
H. Golchin and M. R. Mehdizadeh, Quasi-cosmological Traversable Wormholes in f(R) Gravity, The European Physical Journal C volume 79, (2019)

\bibitem{mishra and sharma 2021}
A. K. Mishra, U. K. Sharma, A new shape function for wormholes in f(R) gravity and general relativity, New Astronomy, 88, (2021)

\bibitem{wormholes with constant and variable redshift functions}
N. Godania, G. C. Samanta, Traversable wormholes in f(R) gravity with constant and variable redshift functions, New Astronomy 80 (2020).

\bibitem{Bronnikov:2010tt}
K.~A.~Bronnikov, M.~V.~Skvortsova and A.~A.~Starobinsky,
Grav. Cosmol. \textbf{16}, 216-222 (2010)
doi:10.1134/S0202289310030047


\bibitem{Antonio-Tsujikawa} A.D. Felice and S. Tsujikawa``f (R) Theories,''

\bibitem{Amendola Dark Energy}
L. Amendola and S.Tsujikawa, Dark Energy: Theory and Observation, Cambridge University Press, (2010)

\bibitem{Bronnikov:2009tv}
K.~A.~Bronnikov and A.~A.~Starobinsky,
Mod. Phys. Lett. A \textbf{24}, 1559-1564 (2009)
doi:10.1142/S0217732309030928

\bibitem{Dolgove fRR}
A. D. Dolgov and M. Kawasaki, Can modified gravity explain accelerated cosmic
expansion?, Phys. Lett. B 573, 1 (2003).
\bibitem{Cosmic  perturbations in f(R) gravity}
R. Bean, D. Bernat, L. Pogosian, A. Silvestri, and M. Trodden, Dynamics of linear perturbations in $f(R)$ gravity, Phys. Rev. D 75 (2007), 064020.

\bibitem{star07}
A.~A.~Starobinsky,
JETP Lett.\  {\bf 86}, 157 (2007).

\bibitem{Sebastian} B. Bahamonde, M. Jamil, P. Pavlovic and Marko Sossich Phys. Rev. D 94, (2016).


\bibitem{Lobo:2009ip} F.~S.~N.~Lobo and M.~A.~Oliveira,
``Wormhole geometries in f(R) modified theories of gravity,''  Phys.\ Rev.\ D {\bf 80}, 104012 (2009)	

\bibitem{Hawking} S. W. Hawking and G. F. R. Ellis Cambridge University Press, (1973).


  \bibitem{Cognola:2007zu} 
G.~Cognola, E.~Elizalde, S.~Nojiri, S.~D.~Odintsov, L.~Sebastiani and S.~Zerbini,
``A Class of viable modified f(R) gravities describing inflation and the onset of accelerated expansion,''
Phys.\ Rev.\ D {\bf 77}, 046009 (2008)


``Non-singular exponential gravity: a simple theory for early- and late-time accelerated expansion,''
Phys.\ Rev.\ D {\bf 83}, 086006 (2011)



\bibitem{Elizalde:2010ts} 
E.~Elizalde, S.~Nojiri, S.~D.~Odintsov, L.~Sebastiani and S.~Zerbini,

\bibitem{Tsujikawa:2007xu} 
S.~Tsujikawa,
Phys.\ Rev.\ D {\bf 77}, 023507 (2008)
[arXiv:0709.1391 [astro-ph]].

\bibitem{HuSa Equation}
W.~Hu and I.~Sawicki,
Phys.\ Rev.\  D {\bf 76}, 064004 (2007).
\bibitem{Louis Yang} C. C. Lee, C. Q. Geng and L. yang,  Singularity phenomena in viable f(R) gravity,Progress of Theoretical Physics, Volume 128, 415-427 (2012).
	

\bibitem{Geng}
C. Q. Geng, Y. T. Hsu and J. R. Lu, Cosmological Constraints on Non-flat Exponential f(R) Gravity, The Astrophysical Journal 926, 74 (2022).

\bibitem{S. Capozziello 2007}
S. Capozziello, V. F. Cardone and A. Troisi, Low surface brightness galaxies rotation curves in the low energy limit of $R^{n}$ gravity : no need for dark matter?, Mon.Not.Roy.Astron.Soc, 375, (2007).

\bibitem{Ali}
A. Ali, R. Gannouji, M. Sami, and A. A. Sen,Background cosmological dynamics in $f(R)$ gravity and observational constraints, Phys. Rev. D 81, 104029 (2010)

\bibitem{lagf} W. Rindler, Relativity, Special, General and Cosmology (Oxford University Press, New York, 2001).	

\bibitem{deflection angle}
S. Weinberg, ‘‘Gravitation and cosmology: principles and applications of the general theory of
relativity,’’ Wiley (1972).

\bibitem{Bronnikov:2006pt}
K.~A.~Bronnikov and A.~A.~Starobinsky,
JETP Lett. \textbf{85}, 1-5 (2007)
doi:10.1134/S0021364007010018

\end{thebibliography}
\end{document}